\documentclass[sigconf]{acmart}
\AtBeginDocument{%
  }

\setcopyright{acmlicensed}
\copyrightyear{2026}
\acmYear{2026}
\acmDOI{XXXXXXX.XXXXXXX}
\acmConference[KDD '26]{Proceedings of the 32nd ACM SIGKDD Conference on Knowledge Discovery and Data Mining}
{August, 2026}{Jeju, Korea}

\acmISBN{978-1-4503-XXXX-X/2018/06}

\usepackage{tabularx}   
\usepackage{geometry}   
\usepackage{graphicx}
\usepackage{subcaption} 
\usepackage{booktabs} 
\usepackage{multirow} 
\usepackage{hyperref}
\usepackage{xcolor}
\usepackage{pifont}   

\newcommand{\xmark}{\textcolor{red}{\ding{55}}}
\newcommand{\cmark}{\textcolor{teal}{\ding{51}}}

\begin{document}

\title{NMRGym: A Comprehensive Benchmark for Nuclear Magnetic Resonance Based Molecular Structure Elucidation}


\author{Zheng Fang}
\email{zfang723@connect.hkust-gz.edu.cn}
\orcid{0009-0009-9735-4350}
\affiliation{%
  \institution{The Hong Kong University of Science and Technology (Guangzhou)}
  \state{Guangzhou}
  \country{China}
}

\author{Chen Yang}
\email{chenyangmiles@gmail.com}
\orcid{}
\affiliation{%
  \institution{The Hong Kong University of Science and Technology (Guangzhou)}
  \state{Guangzhou}
  \country{China}
}

\author{Haitao Yu}
\email{hyu382@connect.hkust-gz.edu.cn}
\orcid{}
\affiliation{%
  \institution{The Hong Kong University of Science and Technology (Guangzhou)}
  \state{Guangzhou}
  \country{China}
}

\author{Haoming Luo}
\email{wangnan8008@gmail.com}
\orcid{}
\affiliation{%
  \institution{Tongji University}
  \state{ShangHai}
  \country{China}
}

\author{Haitao He}
\email{hehaitao@mail.nwpu.edu.cn}
\orcid{}
\affiliation{%
  \institution{Northwestern Polytechnical University}
  \state{Shanxi}
  \country{China}
}

\author{Jiaqing Xie}
\email{xiejiaqing@pjlab.org.cn}
\affiliation{%
  \institution{Shanghai Artificial Intelligence Laboratory}
  \city{Shanghai}
  \country{China}
}

\author{Zhuo Yang}
\email{yangzhuo@pjlab.org.cn}
\affiliation{%
  \institution{Shanghai Artificial Intelligence Laboratory}
  \city{Shanghai}
  \country{China}
}

\author{Yuqiang Li}
\email{liyuqiang@pjlab.org.cn}
\affiliation{%
  \institution{Shanghai Artificial Intelligence Laboratory}
  \city{Shanghai}
  \country{China}
}

\author{Jun Xia}
\email{junxia@hkust-gz.edu.cn}
\orcid{}
\affiliation{%
  \institution{The Hong Kong University of Science and Technology (Guangzhou)}
  \state{Guangzhou}
  \country{China}
}

\renewcommand{\shortauthors}{Zheng et al.}


\begin{abstract}
Nuclear Magnetic Resonance (NMR) spectroscopy is the cornerstone of small-molecule structure elucidation. While deep learning has demonstrated significant potential in automating structure elucidation and spectral simulation, current progress is severely impeded by the reliance on synthetic datasets, which introduces significant domain shifts when applied to real-world experimental spectra. Furthermore, the lack of standardized evaluation protocols and rigorous data splitting strategies frequently leads to unfair comparisons and data leakage. To address these challenges, we introduce \textbf{NMRGym}, the largest and most comprehensive standardized dataset and benchmark derived from high-quality experimental NMR data to date. Comprising \textbf{269,999} unique molecules paired with high-fidelity $^1$H and $^{13}$C spectra, NMRGym bridges the critical gap between synthetic approximations and real-world diversity. We implement a strict quality control pipeline and unify data formats to ensure fair comparison. To strictly prevent data leakage, we enforce a scaffold-based split. Additionally, we provide fine-grained peak-atom level annotations to support future usage. Leveraging this resource, we establish a comprehensive evaluation suite covering diverse downstream tasks, including structure elucidation, functional group prediction from NMR, toxicity prediction from NMR, and spectral simulation, benchmarking representative state-of-the-art methodologies. Finally, we release an open-source leadboard with an automated leaderboard to foster community collaboration and standardize future research. The dataset, benchmark and leaderboard are publicly available at \textcolor{blue}{\url{https://AIMS-Lab-HKUSTGZ.github.io/NMRGym/}}.
\end{abstract}




\begin{teaserfigure}
    \centering
    \includegraphics[width=0.8\textwidth]{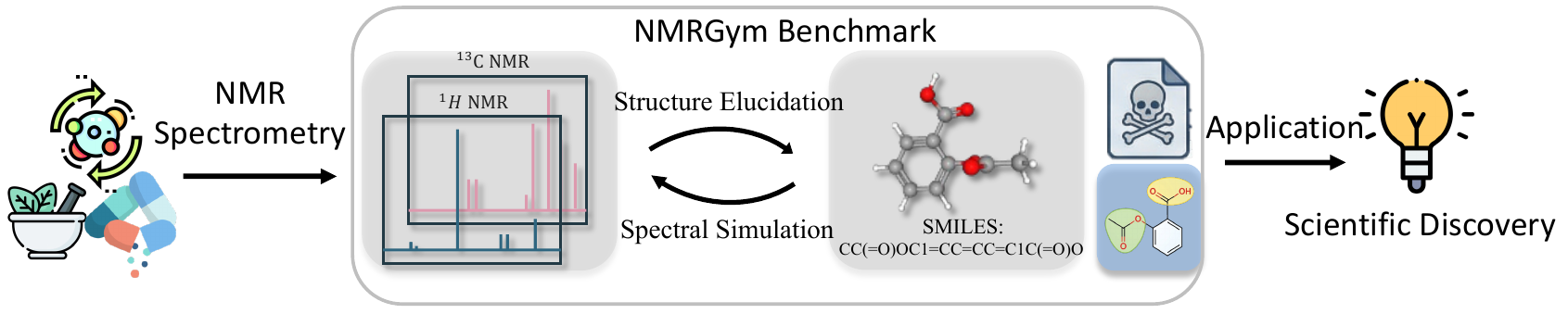}
    \caption{The NMRGym Framework. A comprehensive dataset and benchmark bridging experimental NMR spectrometry and molecular structures to facilitate structure elucidation and spectral simulation—for accelerating scientific discovery.}
    \Description{figure description}
\end{teaserfigure}

\maketitle

\section{Introduction}

Nuclear Magnetic Resonance (NMR) spectroscopy has long been regarded as a cornerstone of small-molecule structure elucidation, providing detailed insights into connectivity, stereochemistry, and functional-group environments that form the basis of modern organic analysis ~\cite{ning2011interpretation,field2012organic}. 
One-dimensional (1D) NMR techniques, particularly proton ($^1$H) and carbon ($^{13}$C) spectroscopy, remain the most widely employed tools in analytical chemistry for molecular \textbf{structure elucidation}\cite{jonas2022prediction}.
Traditional workflows for NMR structure elucidation synergistically integrate empirical rules, physics-based predictions, and database matching, where chemists initially interpret established chemical-shift trends and coupling patterns to assemble local molecular fragments ~\cite{ning2011interpretation,field2012organic}. 
To enhance the reliability of this process, physics-based methods such as Density Functional Theory (DFT) ~\cite{dft} are frequently employed for \textbf{spectral simulation}. These simulations serve as  two critical roles: augmenting reference databases with theoretical data and verifying candidate structures by rigorously comparing predicted spectra against experimental observations\cite{lodewyk2012computational}. 
However, despite their precision, the prohibitive computational cost of DFT restricts its routine application for large-scale database expansion ~\cite{ditchfield1974self,wolinski1990efficient}. 
Consequently, database-driven tools like NMRShiftDB ~\cite{steinbeck2003nmrshiftdb} rely heavily on the limited availability of high-quality experimental NMR data, which remains expensive and time-consuming to acquire ~\cite{review_guo2025artificial}. These limitations collectively highlight the urgent need for more scalable, data-driven approaches.

Motivated by the transformative impact of deep learning in molecular sciences~\cite{jumper2021highly}, recent research in NMR analysis has focused on two pivotal challenges: \textbf{\textit{structure elucidation}} and \textbf{\textit{spectral simulation}}. 
For structure elucidation, methodologies have rapidly evolved from traditional database-driven search~\cite{jin2025nmr} to generative deep learning paradigms. These include Transformer-based sequence translation models~\cite{tan2025clams,hu2024accurate,xue2025nmrmind} that interpret spectral patterns, and emerging diffusion-based approaches~\cite{yang2025diffnmr,xiong2025atomic} capable of generating molecular graphs or 3D conformations. 
Concurrently, forward spectral simulation has progressed from relying on fixed vector descriptors~\cite{binev2007prediction} to sophisticated 2D~\cite{jonas2019rapid, kang2020predictive, kwon2020neural}and 3D~\cite{jonas2019rapid,nmrnet,detanet} geometric representations, enabling more accurate modeling of chemical environments.
Despite the progress, the field fundamentally lacks a standardized benchmark derived from high-quality experimental data. This deficiency creates three critical impediments to future development:
\textbf{1. Domain Shift.} Most current methodologies are validated primarily on synthetic datasets ~\cite{QM9-NMR, hu2024accurate, nmrformer}. Consequently, models trained on such idealized data often encounter severe performance degradation when applied to real-world experimental spectra ~\cite{xiong2025atomic}.
\textbf{2. Inconsistent Evaluation Protocols.} The heterogeneity of spectral metadata—ranging from detailed peak attributes to sparse chemical shift lists—forces models to rely on diverse input formats. This lack of standardization not only limits the broad generalization of models but also precludes fair, quantitative comparisons across different methods.
\textbf{3. Risk of Data Leakage.} The absence of rigorous splitting protocols frequently allows structurally analogous molecules to overlap between training and test sets\cite{wu2018moleculenet} . This necessitates the adoption of scaffold splitting to prevent inflated metrics and accurately evaluate true generalization. 
While experimental datasets like NMRShiftDB ~\cite{steinbeck2003nmrshiftdb} exist, their utility is severely limited by data size, which fails to satisfy the data-hungry requirements of modern deep learning methodologies.

To address these issues, we introduce \textbf{NMRGym}, the most comprehensive and standardized experimental NMR dataset and benchmark to date. 
Through a rigorous quality control (QC) pipeline designed to unify disparate spectral formats, we curate \textbf{269,999} high-fidelity experimental molecule-spectrum pairs, offering a robust resource that enables the rigorous assessment of models within complex, real-world experimental scenarios. Finally, we adopt a scaffold-based split to strictly prevent data leakage and utilize this rigorous framework to conduct a systematic benchmark of representative state-of-the-art methodologies {covering four downstream tasks}, establishing reliable baselines for future research.

Our primary contributions are summarized as follows:
\vspace{-1.5mm}
\begin{itemize}
    \item \textbf{Large-Scale Experimental Data.} We release the largest standardized experimental NMR dataset comprising 269,999 molecules. By providing high-fidelity real-world spectra rather than synthetic approximations, this resource effectively mitigates the domain gap, satisfying the data-hungry requirements of modern deep learning models.
    \item \textbf{Rigorous and Fair Benchmark.} We establish a comprehensive evaluation suite encompassing four downstream tasks (structure elucidation, functional group prediction from NMR, toxicity prediction from NMR, and spectral simulation). Crucially, we introduce scaffold-based splitting to assess out-of-distribution (OOD) robustness and provide fine-grained peak-atom level annotations, facilitating the development of physically interpretative models.
    \item \textbf{Interactive Evaluation Leadboard.} We develop an open-source leadboard featuring an automated leaderboard and visualization tools. This leadboard streamlines the model submission and evaluation process, fostering community collaboration and enabling direct, transparent comparisons of SOTA methodologies.
\end{itemize}

\begin{table}[t]
\centering
\caption{Summary of commonly used NMR datasets. ``\textbf{Exp.}'' denotes whether spectra are from real-world experiments. ``\textbf{Assign.}'' indicates if peak-level assignments are provided.}
\vspace{-3mm}
\label{dataset}
\resizebox{0.9\linewidth}{!}{
\begin{tabular}{l c c c c}
\toprule
\textbf{Dataset} &
\textbf{\# Mols} &
\textbf{Split} &
\textbf{Exp.} &
\textbf{Assign.} \\ 
\midrule
\multicolumn{5}{l}{\textit{DFT Simulation Dataset}} \\
\midrule

QM9-NMR ~\cite{QM9-NMR}
& $\sim 131\,\mathrm{k}$ & Rand. & \xmark & \xmark \\

SpectraBase-Mnova ~\cite{hu2024accurate}
& $\sim 143\,\mathrm{k}$ & Rand. & \xmark & \xmark \\

Multispec ~\cite{multispec}
& $\sim 790\,\mathrm{k}$ & N.A. & \xmark & \xmark \\

Pistachio-Mnova ~\cite{nmrformer}
& $\sim 1.03\,\mathrm{M}$ & N.A. & \xmark & \xmark \\

Mind-Gaussian ~\cite{xue2025nmrmind}
& $\sim 2.25\,\mathrm{M}$ & Rand. & \xmark & \xmark \\

\midrule
\multicolumn{5}{l}{\textit{ML Simulation Dataset}} \\
\midrule

NN-NMR ~\cite{NN-NMR}
& $16\,\mathrm{k}$ & Rand. & \xmark & \xmark \\

ShiftML ~\cite{shiftML}
& $\sim 257\,\mathrm{k}$ & Rand. & \xmark & \xmark \\

Pubchem-NMRNet ~\cite{nmrnet}
& $\sim 106\,\mathrm{M}$ & N.A. & \xmark & \cmark \\

\midrule
\multicolumn{5}{l}{\textit{Experimental Dataset}} \\
\midrule

NMRShiftDB ~\cite{steinbeck2003nmrshiftdb} 
& 43{,}580 & Rand. & \cmark & \cmark \\

NMRBank ~\cite{NMRBank}
& 149{,}135 & N.A. & \cmark & \xmark \\

\midrule

NMRGym (Ours) 
& \textbf{269{,}999} & Scaffold & \cmark & \cmark \\

\bottomrule
\end{tabular}
}
\end{table}

















\vspace{-2.5mm}
\section{Related Work}

\subsection{NMR Structure Elucidation}
The primary objective of structure elucidation is the inverse mapping from experimental NMR spectra to molecular structures. 
Early data-driven approaches reframed structure elucidation as a sequence-to-sequence translation task. A key differentiator among these models is their spectral encoding strategy. For instance, CLAMS ~\cite{tan2025clams} processes NMR spectra as 2D images, utilizing CNNs for visual feature extraction. In contrast, NMRFormer ~\cite{nmrformer} employs a 1D-CNN architecture to automatically learn spectral tokenization. Moving toward more explicit representations, models such as NMR2Struct~\cite{hu2024accurate} and NMRMind ~\cite{xue2025nmrmind} manually tokenize spectra by encoding specific chemical shift ranges and their corresponding signal intensities, allowing transformers to attend to discrete spectral inputs.

To overcome the issues of chemical invalidity often found in autoregressive decoding, search-based frameworks like NMR-Solver~\cite{jin2025nmr} have been introduced. These methods integrate neural representations with rigorous chemical-constrained priors and leverage large-scale simulated databases, such as NMRNet~\cite{nmrnet}, to perform similarity-based retrieval and structural optimization. Recently, inspired by breakthroughs in generative modeling~\cite{hoogeboom2022equivariant, morehead2024geometry, liu2025next, wang2023swallowing}, the field has shifted toward diffusion-based paradigms. DiffNMR ~\cite{yang2025diffnmr} implements a diffusion framework to progressively denoise and generate molecular 2D graphs. Meanwhile, ChefNMR ~\cite{xiong2025atomic} directly generates 3D atomic conformations, capturing the inherent geometric symmetries of molecular structures.
\vspace{-3mm}
\subsection{NMR Spectral Simulation}
Spectral simulation, or the forward problem, focuses on predicting NMR spectra directly from molecular structures to bypass the high computational cost of quantum mechanical calculations. 
The efficacy of these simulators is largely determined by their molecular representation. Early efforts utilized {vector-based representations} ~\cite{binev2007prediction, gerrard2020impression, lin2022machine}, which map fixed-length descriptors to chemical shifts or coupling constants. To better reflect the topological connectivity of organic molecules, researchers transitioned to {2D graph-based representations} ~\cite{jonas2019rapid, kang2020predictive, kwon2020neural}. These models utilize GNNs or Message Passing Neural Networks (MPNN) to capture the local chemical environment of each individual atom. The state-of-the-art has recently moved toward {3D molecular representations} ~\cite{guan2021real, nmrnet, detanet}, incorporating explicit atomic coordinates. By utilizing 3D-aware GNNs, these models can account for long-range through-space interactions that are critical for achieving DFT-level accuracy in chemical shift prediction.
\vspace{-3mm}
\subsection{NMR Data and Benchmarking}
Despite the architectural advancements in NMR machine learning, the field remains significantly hindered by the scarcity, heterogeneity, and lack of standardization in available datasets. As summarized in Table~\ref{dataset}, current research typically relies on two data sources: synthetic datasets and experimental datasets.
Synthetic datasets generated via DFT or ML-based simulators, such as QM9-NMR ~\cite{QM9-NMR}, Multispec ~\cite{multispec}, and Pubchem-NMRNet ~\cite{nmrnet} containing over 100 million entries, provide the scale necessary to train deep neural networks. However, these simulated spectra often fail to capture the nuances of real-world acquisition, such as solvent effects, baseline noise, and complex coupling patterns. Consequently, models trained on these synthetic priors frequently encounter a substantial domain shift, leading to poor generalization when deployed on experimental samples ~\cite{xiong2025atomic}. Conversely, experimental datasets like NMRShiftDB ~\cite{steinbeck2003nmrshiftdb} offers the most authentic representation of chemical reality but are often restricted in size.



\begin{figure*}[t]
    \centering
    \includegraphics[width=0.8\linewidth]{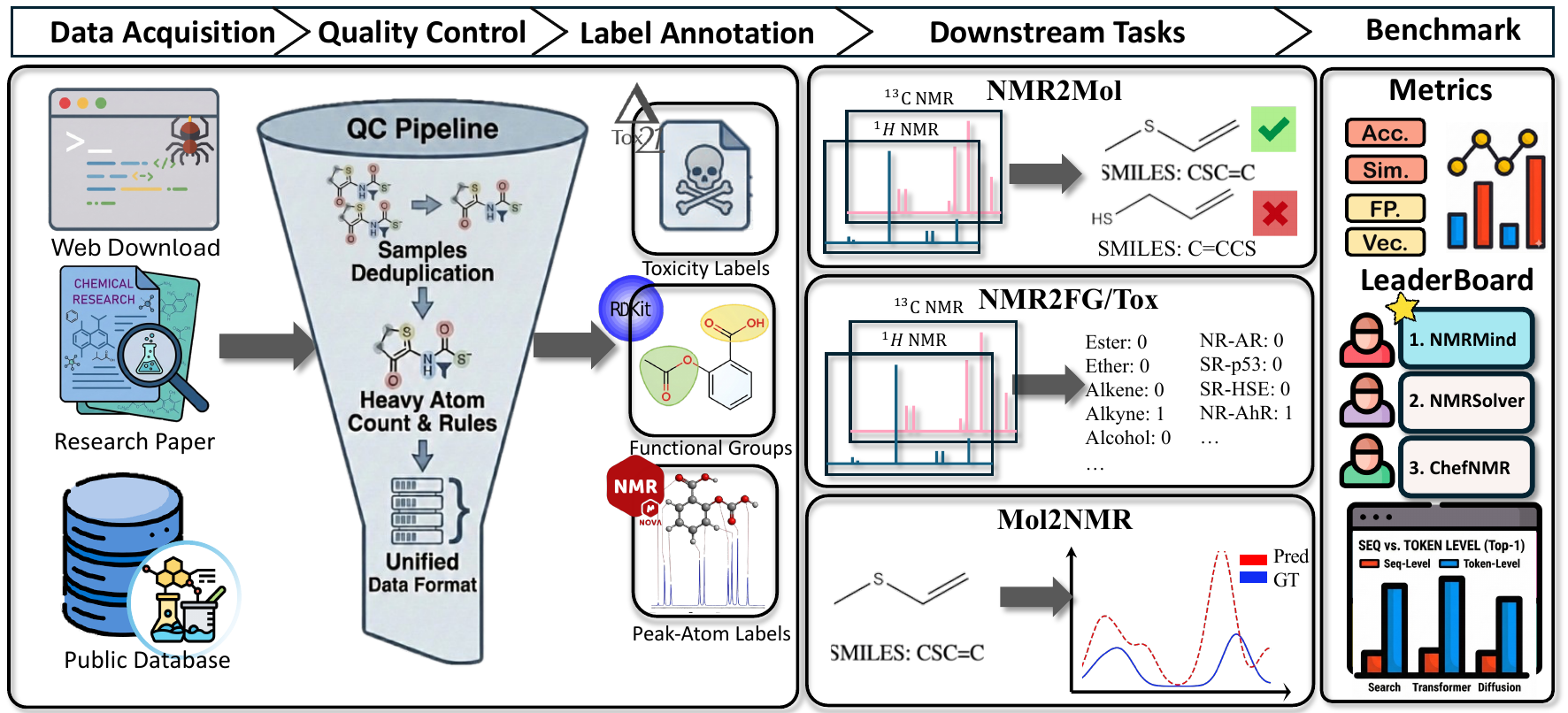}
    \vspace{-2.5mm}
    \caption{Overview of the NMRGym data curation workflow. The pipeline integrates data acquisition from heterogeneous sources, rigorous quality control for standardization, and comprehensive label annotation (including toxicity, functional groups, and peak-atom assignments) to support downstream generative and predictive tasks.}
    \label{method}
\end{figure*}

\vspace{-2.5mm}
\section{Methods}

\subsection{Preliminaries}

We formally define an NMR spectrum $\mathcal{S}$ as an unordered set of $N$ peaks, denoted as $\mathcal{S} = \{s_1, s_2, \dots, s_N\}$, where each $s_i \in \mathbb{R}$ represents the chemical shift value (in ppm). Detailed chemical definitions are provided in Appendix~\ref{chemical}. Since the physical measurement of NMR peaks is invariant to permutation, we treat $\mathcal{S}$ strictly as a set rather than a sequence. The corresponding molecular structure is represented as a SMILES ~\cite{weininger1988smiles} sequence $Y = \{y_1, \dots, y_L\}$. Finally, an overview of the proposed dataset and benchmark construction pipeline is illustrated in Figure~\ref{method}.
\vspace{-3mm}
\subsection{Data Collection and Pre-processing}
\textbf{Data Acquisition.}
To construct a comprehensive and diverse benchmark, we aggregated experimental NMR spectra from a wide array of public chemical databases. Our primary data sources include  CH-NMR-NP ~\cite{asakura2015chnmrnp}, CASCADE ~\cite{guan2021cascade}, SDBS ~\cite{saito2011sdbs}, PubChem ~\cite{kim2025pubchem}, NMRShiftDB ~\cite{steinbeck2003nmrshiftdb}, HMDB ~\cite{wishart2022hmdb}, NP-MRD ~\cite{wishart2022npmrd}, and NMRMind ~\cite{xue2025nmrmind}. The data acquisition process involved a hybrid approach utilizing public APIs, bulk database dumps, and literature mining scripts inspired by the methodologies in ~\cite{xue2025nmrmind}. This multi-source strategy ensures the dataset covers a broad chemical space, ranging from simple organic molecules to complex natural products.

\noindent\textbf{Quality Control (QC).}
We employed RDKit\cite{landrum2013rdkit} to generate canonical SMILES\cite{weininger1988smiles} for all molecules, serving as the unique identifier to detect and remove duplicate entries across different sources. Guided by prior research suggesting that the synergistic use of carbon and proton spectra significantly enhances model performance ~\cite{alberts2023learning, xue2025nmrmind}, we strictly curated the dataset to retain only samples possessing paired $^{1}$H and $^{13}$C NMR spectra.
To ensure the dataset's relevance to organic chemistry research, we filtered molecules based on elemental composition, retaining only those containing heavy atoms from the set $\{C, H, O, N, F, P, S, Cl, Br, I\}$. Furthermore, unlike standard small-molecule datasets that often restrict sequence length, we deliberately retained complex natural products (specifically from NP-MRD dataset) with longer isomeric sequences. This inclusion is critical for evaluating the benchmark's generalization capabilities on heavy-atom structures and macrocyclic compounds.
%

\noindent\textbf{Data Format Standardization.} Unifying NMR data formats from heterogeneous sources presents significant challenges due to inconsistent reporting standards. We addressed two primary inconsistencies to ensure rigorous data uniformity. First, regarding J-coupling constants, we excluded this attribute not only due to its sparsity across public datasets but also because current model architectures are generally not designed to accommodate this specific modality. Second, we addressed peak intensity, which is conventionally utilized in $^{1}$H NMR to deduce proton stoichiometry via signal integration. Instead of forcing the model to learn the complex mapping from inconsistent raw intensity values to discrete atom counts, we adopted a \textit{multiplicity-based encoding strategy}. Specifically, intensity is implicitly represented by the frequency of chemical shift occurrences; for instance, a signal corresponding to two protons (2H) is tokenized by duplicating its chemical shift value twice in the input sequence. This approach explicitly embeds proton count constraints into the unified, sequence-based format, effectively relieving the model from the burden of decoding stoichiometric information from variable intensity inputs.

\noindent\textbf{Label Annotations.} 
Beyond spectrum-SMILES pairs, NMRGym provides diverse annotations to support diverse downstream tasks.

\noindent\textit{1. Functional Groups.} 
Following the protocol established in NMRFormer ~\cite{nmrformer}, we employed RDKit ~\cite{landrum2013rdkit} to identify functional groups via SMARTS substructure matching. The comprehensive definitions of the tracked functional groups are detailed in Appendix \ref{func_def}.

\noindent\textit{2. Toxicity.} 
To demonstrate the dataset's utility in real-world safety assessment, we incorporated toxicity labels from the Tox21 ~\cite{richard2020tox21}. Due to the limited scale of the original Tox21 dataset, we release a targeted subset specifically designed to benchmark molecular toxicity prediction in practical applications. The detailed definitions of the 12 toxicity assays are provided in Appendix \ref{tox_def}.

\noindent\textit{3. Peak-Atom Assignments.} To facilitate interpretability and structure verification —a critical component of the elucidation workflow—we utilized Mnova ~\cite{mnova} software to generate peak-atom level assignments. By providing these detailed annotations, we aim to support future research focused on interpretable verification.

\noindent\textbf{Scaffold Split.} To rigorously evaluate the generalization capability of models on unseen chemical structures, we purposefully avoided random splitting, which results in structural redundancy between training and test sets. Instead, we adopted a scaffold-splitting strategy inspired by recent benchmarking protocols ~\cite{bushuiev2024massspecgym}. We utilized the Bemis-Murcko scaffold decomposition algorithm provided by DeepChem ~\cite{deepchem} to cluster molecules based on their core structural frameworks. Molecules sharing the same scaffold were strictly assigned to the same subset. This strategy ensures that the test set consists of molecules with molecular backbones entirely distinct from those seen during training, enforcing a rigorous evaluation of OOD generalization. The detailed analysis is shown in Sec~\ref{main}.

\begin{figure}[t]
    \centering
    \begin{subfigure}[b]{0.49\textwidth}
        \centering
        \includegraphics[width=\linewidth]{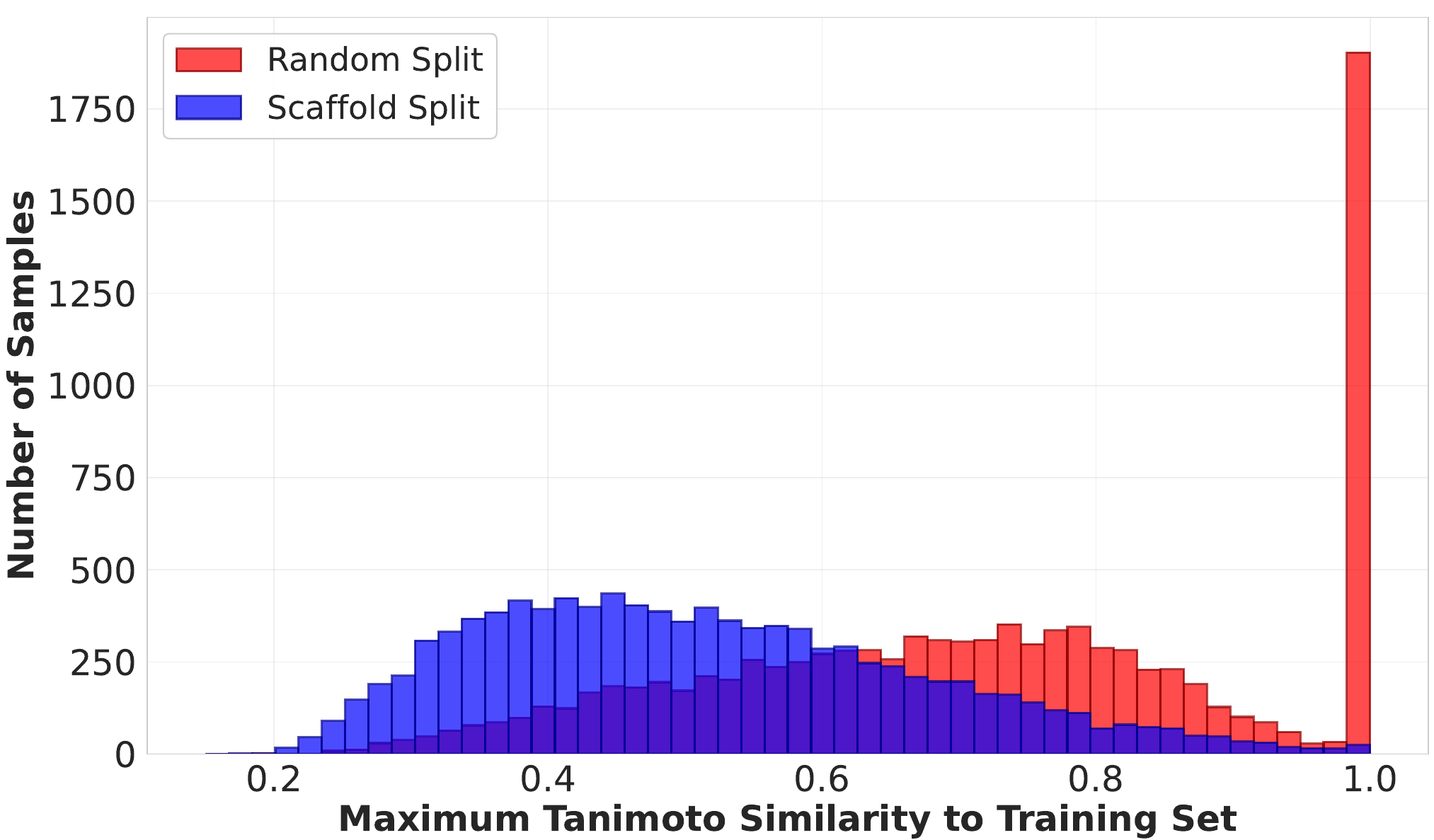}
        \caption{Similarity Distribution}
        \label{fig:sim_dist}
    \end{subfigure}
    \hfill 
    \begin{subfigure}[b]{0.49\textwidth}
        \centering
        \includegraphics[width=\linewidth]{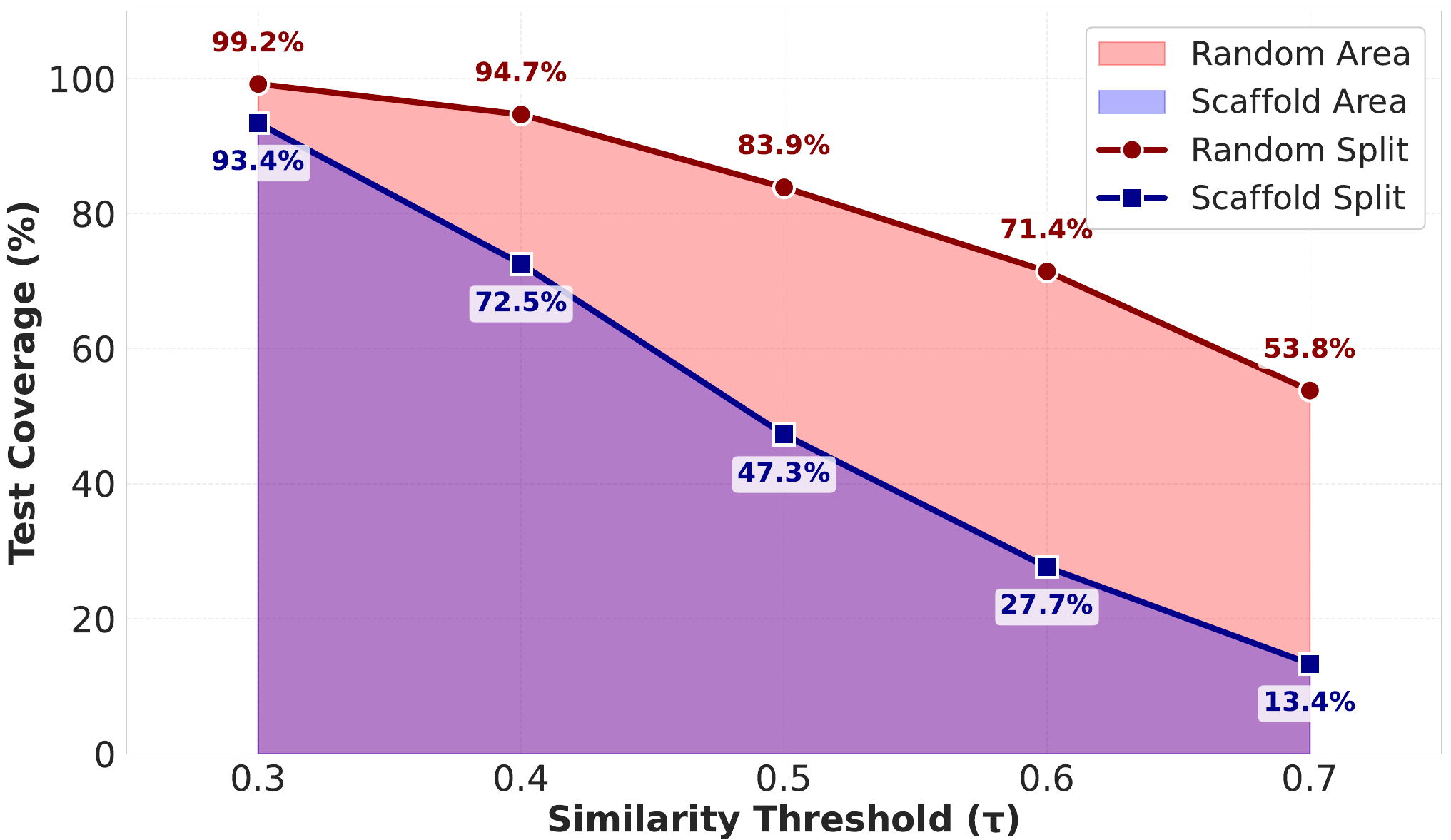}
        \caption{Coverage@$\tau$ Curve}
        \label{fig:cov_curve}
    \end{subfigure}
    \vspace{-6mm}
    \caption{\textbf{Comparison of data leakage between Random (red) and Scaffold (blue) splits.} 
    (a) Distribution of Maximum Tanimoto Similarity between test and training sets. The scaffold split shows a distinct shift towards lower similarity. 
    (b) Coverage@$\tau$ curves measuring the fraction of test molecules with structural neighbors in the training set ($\text{similarity} \ge \tau$).}
    \label{fig:split_analysis}
\end{figure}

\subsection{Task Formulation and Evaluation}


\begin{table*}[t]
\centering
\caption{Comparison of Sequence-Level and Token-Level Structure Elucidation Performance Across Methods (NMR2Mol). \textit{Formula Condition} denotes the utilization of the molecular formula as a structural constraint.}
\vspace{-3mm}
\label{nmr2smiles1}
\setlength{\tabcolsep}{11pt}
\begin{tabular}{lcccccc}
\toprule
 & \multicolumn{2}{c}{\textbf{Top-1 (\% ↑)}} & \multicolumn{2}{c}{\textbf{Top-5 (\% ↑)}} & \multicolumn{2}{c}{\textbf{Top-10 (\% ↑)}} \\
\cmidrule(lr){2-3} \cmidrule(lr){4-5} \cmidrule(lr){6-7}
\textbf{Model} & Seq-Level & Token-Level & Seq-Level & Token-Level & Seq-Level & Token-Level \\
\midrule
\multicolumn{7}{l}{\textit{Search-based Methods}} \\
NMR-Solver(Search-Only)\cite{jin2025nmr} 
& 6.27 {\scriptsize $\pm$ 0.00} 
& 21.75 {\scriptsize $\pm$ 0.00} 
& 13.96 {\scriptsize $\pm$ 0.00} 
& 34.78 {\scriptsize $\pm$ 0.00} 
& 17.37 {\scriptsize $\pm$ 0.00} 
& 40.13 {\scriptsize $\pm$ 0.00}  \\ 

{\quad\itshape +Formula Condition}
& \textbf{17.92 {\scriptsize $\pm$ 0.00}} 
& 27.64  {\scriptsize $\pm$ 0.00} 
& \textbf{33.97 {\scriptsize $\pm$ 0.00}} 
& 41.06 {\scriptsize $\pm$ 0.00} 
& \textbf{36.48 {\scriptsize $\pm$ 0.00}} 
& 42.79 {\scriptsize $\pm$ 0.00} \\ 

\midrule

\multicolumn{7}{l}{\textit{Transformer-based Methods}} \\

CLAMS\cite{tan2025clams}
& 0.00 {\scriptsize $\pm$ 0.00}
& 7.90 {\scriptsize $\pm$ 0.07}
& 0.00 {\scriptsize $\pm$ 0.00}
& 12.78 {\scriptsize $\pm$ 0.13}
& 0.00 {\scriptsize $\pm$ 0.00}
& 14.70 {\scriptsize $\pm$ 0.14} \\

NMRFormer\cite{nmrformer} 
& 1.75 {\scriptsize $\pm$ 0.04}
& 22.49 {\scriptsize $\pm$ 0.02}
& 2.81 {\scriptsize $\pm$ 0.05}
& 29.58 {\scriptsize $\pm$ 0.02}
& 3.30 {\scriptsize $\pm$ 0.03}
& 32.32 {\scriptsize $\pm$ 0.01} \\

NMR2Struct\cite{hu2024accurate}
& 0.24 {\scriptsize $\pm$ 0.08} 
& 21.97 {\scriptsize $\pm$ 0.37}  
& 1.05 {\scriptsize $\pm$ 0.19}  & 33.65 {\scriptsize $\pm$ 0.55} & 1.85 {\scriptsize $\pm$ 0.27}  & 38.67 {\scriptsize $\pm$ 0.63} \\

NMRMind\cite{xue2025nmrmind}
& 11.75 {\scriptsize $\pm$ 0.17} 
& 36.96 {\scriptsize $\pm$ 0.01}
& 23.22 {\scriptsize $\pm$ 0.05} 
& 54.71 {\scriptsize $\pm$ 0.00}
& 27.00 {\scriptsize $\pm$ 0.15} 
& 59.08 {\scriptsize $\pm$ 0.01} \\
{\quad\itshape +Formula Condition}
& {15.49 {\scriptsize $\pm$ 0.08}} 
& \textbf{40.82 {\scriptsize $\pm$ 0.06}}
& {29.82 {\scriptsize $\pm$ 0.09}} 
& \textbf{60.00 {\scriptsize $\pm$ 0.00}}
& {34.03 {\scriptsize $\pm$ 0.09}} 
& \textbf{64.40 {\scriptsize $\pm$ 0.11}} \\

\midrule

\multicolumn{7}{l}{\textit{Diffusion-based Methods}} \\
DiffNMR\cite{yang2025diffnmr} 
& 0.00 {\scriptsize $\pm$ 0.00}
& 17.37 {\scriptsize $\pm$ 0.02}
& 0.00 {\scriptsize $\pm$ 0.00}
& 25.37 {\scriptsize $\pm$ 0.01}
& 0.00 {\scriptsize $\pm$ 0.00}
& 28.49 {\scriptsize $\pm$ 0.02} \\

ChefNMR-S\cite{xiong2025atomic}
& 0.02 {\scriptsize $\pm$ 0.00}
& 3.18 {\scriptsize $\pm$ 0.01}
& 0.04 {\scriptsize $\pm$ 0.01}
& 8.93 {\scriptsize $\pm$ 0.00}
& 0.05 {\scriptsize $\pm$ 0.01}
& 11.4 {\scriptsize $\pm$ 0.03} \\

ChefNMR-S(Finetune)\cite{xiong2025atomic}
& 1.69 {\scriptsize $\pm$ 0.02}   
& 11.87 {\scriptsize $\pm$ 0.03}  
& 3.87 {\scriptsize $\pm$ 0.03}  
& 23.85 {\scriptsize $\pm$ 0.07}  
& 5.04 {\scriptsize $\pm$ 0.04}   
& 28.18 {\scriptsize $\pm$ 0.09} \\ 

ChefNMR-L\cite{xiong2025atomic}
& 0.02 {\scriptsize $\pm$ 0.00}
& 3.70 {\scriptsize $\pm$ 0.02}
& 0.07 {\scriptsize $\pm$ 0.01}
& 10.16 {\scriptsize $\pm$ 0.00}
& 0.08 {\scriptsize $\pm$ 0.01}
& 13.29 {\scriptsize $\pm$ 0.02} \\

ChefNMR-L(Finetune)\cite{xiong2025atomic}
& 1.93 {\scriptsize $\pm$ 0.02}
& 13.59 {\scriptsize $\pm$ 0.14}
& 4.36 {\scriptsize $\pm$ 0.08}
& 25.92 {\scriptsize $\pm$ 0.01}
& 5.66 {\scriptsize $\pm$ 0.01}
& 30.07 {\scriptsize $\pm$ 0.02} \\



\bottomrule
\end{tabular}
\end{table*}

\noindent\textbf{Task 1: Structure Elucidation (NMR2Mol).} \\
\textit{Objective.} This task aims to reconstruct the molecular structure, represented as a SMILES sequence $Y = \{y_1, \dots, y_L\}$, from an input spectrum $\mathcal{S}$. The model generally aims to maximize the conditional probability $P(Y|\mathcal{S})$. 
Additionally, for methods that support explicit chemical constraints, we incorporate the {molecular formula} $\mathcal{F}$ as an auxiliary condition. In such cases, the objective is refined to maximize the joint conditional probability $P(Y|\mathcal{S}, \mathcal{F})$.

\noindent\textit{Metrics.} We evaluate the generation quality using three metrics. First, Top-$K$ exact match accuracy measures if the ground-truth SMILES $Y$ exists within the top-$K$ candidates $\{\hat{Y}^{(1)}, \dots, \hat{Y}^{(K)}\}$:
\begin{equation}
    \text{Acc}@K = \frac{1}{M} \sum_{i=1}^{M} \mathbb{I}(Y_i \in \{\hat{Y}_i^{(1)}, \dots, \hat{Y}_i^{(K)}\}).
\end{equation}
Second, token-level accuracy calculates the proportion of correctly predicted tokens at each position relative to the ground truth sequence. Third, we evaluate structural similarity using three distinct molecular fingerprints\cite{landrum2013rdkit}: Morgan, Topological Torsion, and Atom-Pair. All fingerprints are generated as 2048-bit vectors using the RDKit library. Let $\mathbf{v}, \hat{\mathbf{v}} \in \{0, 1\}^{D}$ denote the binary fingerprint vectors of the ground truth and the predicted molecule, respectively. We report the Tanimoto Similarity, defined as the ratio of the intersection to the union of the active bits:
\begin{equation}
    \text{Sim}_{\text{Tanimoto}}(\mathbf{v}, \hat{\mathbf{v}}) = \frac{\mathbf{v} \cdot \hat{\mathbf{v}}}{\|\mathbf{v}\|^2 + \|\hat{\mathbf{v}}\|^2 - \mathbf{v} \cdot \hat{\mathbf{v}}}.
\end{equation}
We also compute the Cosine Similarity to measure the orientation alignment between the high-dimensional fingerprint vectors:
\begin{equation}
    \text{Sim}_{\text{Cosine}}(\mathbf{v}, \hat{\mathbf{v}}) = \frac{\mathbf{v} \cdot \hat{\mathbf{v}}}{\|\mathbf{v}\|_2 \|\hat{\mathbf{v}}\|_2}.
\end{equation}
For Transformer-based models, we employ {beam search} to approximate the optimal sequence by iteratively expanding the most probable partial tokens. 
For diffusion-based models, we utilize a {Top-$K$ sampling} strategy, where a batch of $K$ candidate structures is generated directly via the reverse diffusion process. 
Finally, search-based approaches operate by retrieving the molecule following a {similarity} metric~\cite{jin2025nmr} with the input $\mathcal{S}$.

\noindent\textbf{Task 2: Property Prediction (NMR2FG \& NMR2Tox).} \\
\textit{Objective.} We formulate functional group and toxicity prediction as multi-label binary classification tasks. Given a spectrum $\mathcal{S}$, the goal is to predict a binary label vector $\mathbf{y} \in \{0, 1\}^C$ covering $C$ properties.

\noindent\textit{Metrics.} Following standard benchmarks, we report Macro/Micro-F1, Recall, and Accuracy to evaluate class-wise and global performance, accounting for label imbalance. We also report Subset Accuracy, the strictest metric requiring all $C$ labels for a given sample to be correctly predicted:
\begin{equation}
    \text{Acc}_{subset} = \frac{1}{M} \sum_{i=1}^{M} \mathbb{I}(\mathbf{y}_i = \hat{\mathbf{y}}_i).
\end{equation}

\noindent\textbf{Task 3: Spectral Simulation (Mol2NMR).} \\
\textit{Objective.} This task focuses on the forward prediction of spectral peaks from a molecular structure. Unlike previous methods restricted to specific inputs, we generalize the input as a molecular representation $\mathcal{M}$ (e.g., a 2D molecular graph or a 3D conformer). The objective is to predict the set of chemical shifts $\hat{\mathcal{S}} = \{\hat{s}_1, \dots, \hat{s}_{\hat{N}}\}$ that closely approximates the ground truth spectrum $\mathcal{S}$.

\noindent\textit{Metrics.} To evaluate spectral fidelity, we employ a dual-metric protocol consisting of Vector Similarity for global retrieval and Set Similarity for fine-grained scoring, following the approach outlined in ~\cite{jin2025nmr}. For Vector Similarity, we smooth discrete peaks into continuous vectors using a Gaussian kernel with bandwidths $\sigma_{\text{vec}}$ of 0.3 ppm ($^{1}$H) and 2.0 ppm ($^{13}$C). We then compute the Cosine Similarity to measure the overlap of spectral envelopes:
\begin{equation}
    \text{Sim}_{\text{vec}}(\mathbf{v}, \hat{\mathbf{v}}) = \frac{\mathbf{v} \cdot \hat{\mathbf{v}}}{\|\mathbf{v}\|_2 \|\hat{\mathbf{v}}\|_2}.
\end{equation}
For Set Similarity, we focus on peak-level precision by formulating the comparison as a bipartite matching problem. Using the Hungarian algorithm ~\cite{kuhn1955hungarian}, we find the optimal permutation $\pi$ that minimizes the displacement between predicted and ground truth peaks. The similarity score is normalized by a tolerance $\sigma_{\text{set}}$ (1.0 ppm for $^{1}$H, 10.0 ppm for $^{13}$C):
\begin{equation}
    \text{Sim}_{\text{set}}(\mathcal{S}, \hat{\mathcal{S}}) = \exp\left( - \frac{1}{N} \sum_{i=1}^{N} \frac{| s_i - \hat{s}_{\pi(i)} |}{\sigma_{\text{set}}} \right).
\end{equation}


\begin{table*}[t]
\centering
\caption{Top-K Morgan, torsion, and atom-pair fingerprint similarity evaluation using Tanimoto similarity. (NMR2Mol)}
\vspace{-3mm}
\label{nmr2smiles2}
\resizebox{\linewidth}{!}{
\begin{tabular}{l c c c c c c c c c}
\toprule
& \multicolumn{3}{c}{\textbf{Morgan Sim (\% ↑)}} 
& \multicolumn{3}{c}{\textbf{Torsion Sim (\% ↑)}} 
& \multicolumn{3}{c}{\textbf{Atom-Pair Sim (\% ↑)}} \\
\cmidrule(lr){2-4}
\cmidrule(lr){5-7}
\cmidrule(lr){8-10}
& Top-1 & Top-5 & Top-10
& Top-1 & Top-5 & Top-10
& Top-1 & Top-5 & Top-10 \\
\midrule

\multicolumn{10}{l}{\textit{Search-based Methods}} \\
NMR-Solver(Search-Only)\cite{jin2025nmr}
& 33.28 {\scriptsize $\pm$ 0.00} 
& 42.88 {\scriptsize $\pm$ 0.00}
& 46.84 {\scriptsize $\pm$ 0.00}
& 34.63 {\scriptsize $\pm$ 0.00}
& 45.29 {\scriptsize $\pm$ 0.00}
& 49.50 {\scriptsize $\pm$ 0.00}
& 45.65 {\scriptsize $\pm$ 0.00}
& 55.48 {\scriptsize $\pm$ 0.00}
& 59.17 {\scriptsize $\pm$ 0.00} \\

{\quad\itshape +Formula Condition}
& 43.31 {\scriptsize $\pm$ 0.00} 
& 45.70 {\scriptsize $\pm$ 0.00}
& 46.11 {\scriptsize $\pm$ 0.00}
& 44.00 {\scriptsize $\pm$ 0.00}
& 46.18 {\scriptsize $\pm$ 0.00}
& 46.54 {\scriptsize $\pm$ 0.00}
& 46.00 {\scriptsize $\pm$ 0.00}
& 47.71 {\scriptsize $\pm$ 0.00}
& 48.01 {\scriptsize $\pm$ 0.00} \\

\midrule

\multicolumn{10}{l}{\textit{Transformer-based Methods}} \\

CLAMS\cite{tan2025clams}
& 0.72 {\scriptsize $\pm$ 0.12}
& 2.24 {\scriptsize $\pm$ 0.39}
& 3.49 {\scriptsize $\pm$ 0.73}
& 0.22 {\scriptsize $\pm$ 0.07}
& 0.75 {\scriptsize $\pm$ 0.17}
& 1.20 {\scriptsize $\pm$ 0.26}
& 0.58 {\scriptsize $\pm$ 0.11}
& 2.10 {\scriptsize $\pm$ 0.41}
& 3.51 {\scriptsize $\pm$ 0.80} \\

NMRFormer\cite{nmrformer}
& 27.87 {\scriptsize $\pm$ 0.01}
& 40.94 {\scriptsize $\pm$ 0.06}
& 44.85 {\scriptsize $\pm$ 0.04}
& 29.11 {\scriptsize $\pm$ 0.01}
& 43.52 {\scriptsize $\pm$ 0.07}
& 47.92 {\scriptsize $\pm$ 0.05}
& 33.25 {\scriptsize $\pm$ 0.05}
& 49.44 {\scriptsize $\pm$ 0.04}
& 53.98 {\scriptsize $\pm$ 0.04} \\

NMR2Struct\cite{hu2024accurate}
& 35.18 {\scriptsize $\pm$ 0.71} 
& 43.43 {\scriptsize $\pm$ 0.93}
& 47.47 {\scriptsize $\pm$ 0.91}
& 32.79 {\scriptsize $\pm$ 0.83}
& 42.64 {\scriptsize $\pm$ 0.98}
& 47.28 {\scriptsize $\pm$ 0.97}
& 39.54 {\scriptsize $\pm$ 0.63}
& 48.26 {\scriptsize $\pm$ 0.74}
& 52.16 {\scriptsize $\pm$ 0.71} \\

NMRMind\cite{xue2025nmrmind}
& 61.14 {\scriptsize $\pm$ 0.12} 
& 67.83 {\scriptsize $\pm$ 0.10}
& 70.23 {\scriptsize $\pm$ 0.10}
& 63.47 {\scriptsize $\pm$ 0.09}
& 70.57 {\scriptsize $\pm$ 0.09}
& 73.06 {\scriptsize $\pm$ 0.07}
& 69.56 {\scriptsize $\pm$ 0.07}
& 75.65 {\scriptsize $\pm$ 0.06}
& 77.67 {\scriptsize $\pm$ 0.06} \\

{\quad\itshape +Formula Condition}
& \textbf{65.51 {\scriptsize $\pm$ 0.03}} 
& \textbf{72.80 {\scriptsize $\pm$ 0.10}}
& \textbf{75.07 {\scriptsize $\pm$ 0.08}}
& \textbf{67.94 {\scriptsize $\pm$ 0.03}}
& \textbf{75.63 {\scriptsize $\pm$ 0.06}}
& \textbf{77.99 {\scriptsize $\pm$ 0.07}}
& \textbf{74.86 {\scriptsize $\pm$ 0.01}}
& \textbf{81.15 {\scriptsize $\pm$ 0.04}}
& \textbf{82.92 {\scriptsize $\pm$ 0.03}} \\

\midrule
\multicolumn{10}{l}{\textit{Diffusion-based Methods}} \\
DiffNMR\cite{yang2025diffnmr}
& 9.05  {\scriptsize $\pm$ 0.02}
& 13.33 {\scriptsize $\pm$ 0.01}
& 14.67 {\scriptsize $\pm$ 0.02}
& 6.44  {\scriptsize $\pm$ 0.01}
& 12.22 {\scriptsize $\pm$ 0.02}
& 14.40 {\scriptsize $\pm$ 0.01}
& 20.48 {\scriptsize $\pm$ 0.02}
& 28.95 {\scriptsize $\pm$ 0.01}
& 31.02 {\scriptsize $\pm$ 0.02} \\

ChefNMR-S\cite{xiong2025atomic}
& 2.65  {\scriptsize $\pm$ 0.02}
& 7.04  {\scriptsize $\pm$ 0.02}
& 9.10  {\scriptsize $\pm$ 0.02}
& 2.33  {\scriptsize $\pm$ 0.02}
& 6.85  {\scriptsize $\pm$ 0.03}
& 9.26  {\scriptsize $\pm$ 0.02}
& 6.86  {\scriptsize $\pm$ 0.04}
& 18.15 {\scriptsize $\pm$ 0.05}
& 23.26 {\scriptsize $\pm$ 0.10} \\

ChefNMR-S(Finetune)\cite{xiong2025atomic}
& 17.17 {\scriptsize $\pm$ 0.07} 
& 33.81 {\scriptsize $\pm$ 0.12}
& 39.42 {\scriptsize $\pm$ 0.15}
& 18.72 {\scriptsize $\pm$ 0.08}
& 37.36 {\scriptsize $\pm$ 0.17}
& 43.79 {\scriptsize $\pm$ 0.18}
& 26.18 {\scriptsize $\pm$ 0.09}
& 49.68 {\scriptsize $\pm$ 0.17}
& 56.40 {\scriptsize $\pm$ 0.19} \\

ChefNMR-L\cite{xiong2025atomic}
& 3.14  {\scriptsize $\pm$ 0.02}
& 8.18  {\scriptsize $\pm$ 0.03}
& 10.45 {\scriptsize $\pm$ 0.02}
& 2.82  {\scriptsize $\pm$ 0.02}
& 7.99  {\scriptsize $\pm$ 0.03}
& 10.68 {\scriptsize $\pm$ 0.02}
& 7.98  {\scriptsize $\pm$ 0.01}
& 20.59 {\scriptsize $\pm$ 0.03}
& 26.05 {\scriptsize $\pm$ 0.00} \\

ChefNMR-L(Finetune)\cite{xiong2025atomic}
& 20.29 {\scriptsize $\pm$ 0.12} 
& 37.62 {\scriptsize $\pm$ 0.00}
& 42.91 {\scriptsize $\pm$ 0.04}
& 22.15 {\scriptsize $\pm$ 0.04}
& 41.57 {\scriptsize $\pm$ 0.03}
& 47.55 {\scriptsize $\pm$ 0.04}
& 30.68 {\scriptsize $\pm$ 0.19}
& 53.97 {\scriptsize $\pm$ 0.10}
& 59.72 {\scriptsize $\pm$ 0.07} \\

\bottomrule
\end{tabular}
}
\end{table*}

\section{Experiments}

\begin{table*}[h]
\centering
\caption{Comparison of Multi-Label Functional Group Prediction Models (NMR2Func).}
\vspace{-3mm}
\label{NMR2Func}
\resizebox{\linewidth}{!}{
\begin{tabular}{lccccccc}
\toprule
\textbf{Model} & 
\textbf{Macro@Acc.\textbf{(\%)}} & 
\textbf{Micro@Acc.\textbf{(\%)}} & 
\textbf{Macro@Rec.\textbf{(\%)}} & 
\textbf{Micro@Rec.\textbf{(\%)}} & 
\textbf{Macro@F1\textbf{(\%)}} & 
\textbf{Micro@F1\textbf{(\%)}} & 
\textbf{Acc.\textbf{(\%)}} \\
\midrule
\multicolumn{8}{l}{\textit{Classic Classification Methods}} \\

CNN
& 91.95 {\scriptsize $\pm$ 1.49}
& 91.95 {\scriptsize $\pm$ 1.49}
& 51.45 {\scriptsize $\pm$ 0.00} 
& 82.67 {\scriptsize $\pm$ 0.00} 
& 38.14 {\scriptsize $\pm$ 7.16}
& 77.42 {\scriptsize $\pm$ 1.72}
& 25.37 {\scriptsize $\pm$ 2.10} \\

MLP
& 92.11 {\scriptsize $\pm$ 1.58}
& 92.11 {\scriptsize $\pm$ 1.58}
& 50.63 {\scriptsize $\pm$ 0.00} 
& 81.88 {\scriptsize $\pm$ 0.00} 
& 38.81 {\scriptsize $\pm$ 5.19}
& 77.63 {\scriptsize $\pm$ 2.00}
& 28.18 {\scriptsize $\pm$ 0.71} \\

XGBoost
& \textbf{94.85 {\scriptsize $\pm$ 0.00}} 
& \textbf{94.85 {\scriptsize $\pm$ 0.00}} 
& 45.02 {\scriptsize $\pm$ 0.00}  
& 81.00 {\scriptsize $\pm$ 0.00} 
& \textbf{55.47 {\scriptsize $\pm$ 0.00}}
& \textbf{85.92 {\scriptsize $\pm$ 0.00}} 
& \textbf{45.68 {\scriptsize $\pm$ 0.00}} \\ 

Random Forest
& 93.11 {\scriptsize $\pm$ 0.01}
& 93.11 {\scriptsize $\pm$ 0.01}
& 41.97 {\scriptsize $\pm$ 0.03} 
& 77.72 {\scriptsize $\pm$ 0.05}  
& 48.64 {\scriptsize $\pm$ 0.00}
& 81.39 {\scriptsize $\pm$ 0.03}
& 36.14 {\scriptsize $\pm$ 0.16} \\

\midrule
\multicolumn{8}{l}{\textit{State-of-Art Methods}} \\

CLAMS\cite{tan2025clams} 
& 91.20 {\scriptsize $\pm$ 2.27}
& 91.20 {\scriptsize $\pm$ 2.27}
& \textbf{51.89 {\scriptsize $\pm$ 0.00}} 
& 80.52 {\scriptsize $\pm$ 0.00} 
& 39.37 {\scriptsize $\pm$ 3.39}
& 75.79 {\scriptsize $\pm$ 3.68}
& 21.48 {\scriptsize $\pm$ 6.77} \\

NMR2Struct\cite{hu2024accurate}
& 92.23 {\scriptsize $\pm$ 0.09}
& 92.23 {\scriptsize $\pm$ 0.09}
& 58.62 {\scriptsize $\pm$ 1.12}
& \textbf{83.24 {\scriptsize $\pm$ 0.49}}
& 55.44 {\scriptsize $\pm$ 0.03} 
& 81.69 {\scriptsize $\pm$ 0.05}
& 36.25 {\scriptsize $\pm$ 0.18} \\

\bottomrule
\end{tabular}
}
\end{table*}

\begin{table*}[htbp]
\centering
\caption{Comparison of Toxicity Prediction Models (NMR2Tox).}
\vspace{-3mm}
\label{NMR2Tox}
\resizebox{\linewidth}{!}{
\begin{tabular}{lccccccc}
\toprule
\textbf{Model} & 
\textbf{Macro@Acc.\textbf{(\%)}} & 
\textbf{Micro@Acc.\textbf{(\%)}} & 
\textbf{Macro@Rec.\textbf{(\%)}} & 
\textbf{Micro@Rec.\textbf{(\%)}} & 
\textbf{Macro@F1\textbf{(\%)}} & 
\textbf{Micro@F1\textbf{(\%)}} & 
\textbf{Acc.\textbf{(\%)}} \\

\midrule

MLP
& 75.88 {\scriptsize $\pm$ 6.31}
& 75.88 {\scriptsize $\pm$ 6.31}
& 00.00 {\scriptsize $\pm$ 0.00} 
& 00.00 {\scriptsize $\pm$ 0.00} 
& 18.76 {\scriptsize $\pm$ 0.93}
& 21.55 {\scriptsize $\pm$ 1.88}
& 28.45 {\scriptsize $\pm$ 9.80} \\

CNN
& 84.22 {\scriptsize $\pm$ 0.24}
& 84.22 {\scriptsize $\pm$ 0.24}
& 00.00 {\scriptsize $\pm$ 0.00} 
& 00.00 {\scriptsize $\pm$ 0.00} 
& \textbf{20.58 {\scriptsize $\pm$ 1.31}}
& \textbf{26.42 {\scriptsize $\pm$ 0.70}}
& 32.08 {\scriptsize $\pm$ 0.86} \\

XGBoost
& \textbf{91.91 {\scriptsize $\pm$ 0.05}}
& \textbf{91.91 {\scriptsize $\pm$ 0.05}}
& \textbf{4.76 {\scriptsize $\pm$ 0.31}} 
& \textbf{4.41 {\scriptsize $\pm$ 0.00}} 
& 7.68 {\scriptsize $\pm$ 0.74}
& 7.94 {\scriptsize $\pm$ 0.04}
& 50.77 {\scriptsize $\pm$ 0.52} \\

Random Forest
& 91.77 {\scriptsize $\pm$ 0.08}
& 91.77 {\scriptsize $\pm$ 0.08}
& 2.01 {\scriptsize $\pm$ 0.00} 
& 2.22 {\scriptsize $\pm$ 0.00} 
& 3.22 {\scriptsize $\pm$ 0.13}
& 4.07 {\scriptsize $\pm$ 0.04}
& \textbf{51.19 {\scriptsize $\pm$ 0.52}} \\

\bottomrule
\end{tabular}
}
\end{table*}

\begin{table}[h]
    \centering
    \caption{Performance Comparison of Spectral Simulation Models. $^*$ denotes closed-source commercial software.}
    \label{tab:nmr_comparison_refined}
    \resizebox{\linewidth}{!}{
        \begin{tabular}{l c c c} 
            \toprule
            \textbf{Method} & \textbf{Coverage} & \textbf{Set Sim.} & \textbf{Vec. Sim.} \\
            \midrule
            \multicolumn{4}{l}{\textit{\textbf{$^{1}$H NMR Prediction}}} \\
            \midrule
            Mnova$^{*}$\cite{mnova} & 99.91\% & \textbf{0.8790} & 0.8443 \\ 
            Cascade (DFTNN)\cite{guan2021cascade} & 97.96\% & 0.6899 & 0.8692 \\
            DetaNet\cite{detanet} & 85.59\% & 0.6871 & 0.8632 \\
            NMRNet\cite{nmrnet} & \textbf{100.0\%} & 0.7046 & \textbf{0.8785} \\
            \midrule
            \multicolumn{4}{l}{\textit{\textbf{$^{13}$C NMR Prediction}}} \\
            \midrule
            Mnova$^{*}$\cite{mnova} & 99.41\% & 0.7268 & \textbf{0.8948} \\ 
            Cascade (DFTNN)\cite{guan2021cascade} & 97.96\% & 0.4474 & 0.3303 \\
            Cascade (ExpNN-DFT)\cite{guan2021cascade} & 97.96\% & 0.1466 & 0.1232 \\
            Cascade (ExpNN-FF)\cite{guan2021cascade} & 97.96\% & 0.0330 & 0.0597 \\
            DetaNet\cite{detanet} & 85.59\% & 0.7784 & 0.6417 \\
            NMRNet\cite{nmrnet} & \textbf{100.0\%} & \textbf{0.9080} & 0.8492 \\
            \bottomrule
            \multicolumn{4}{l}{\footnotesize \textit{* Closed-source / Commercial Method}} \\
        \end{tabular}
    }
\end{table}
\subsection{Benchmark}

\textbf{NMR2SMILES.} To ensure a rigorous and fair comparison, we reproduced all baseline methods following their official implementations. Tables~\ref{nmr2smiles1} and Figure~\ref{nmr2smiles2} present the comparative analysis across search-based, transformer-based, and diffusion-based paradigms.
Among transformer-based approaches, {NMRMind} establishes the state-of-the-art for generative models, achieving a Top-1 sequence accuracy of 15.49\% with formula constraints. More importantly, it demonstrates superior structural comprehension, significantly outperforming all baselines in fingerprint similarity metrics (e.g., 65.51\% Top-1 Morgan similarity compared to 43.31\% for the best search-based baseline). We attribute this robustness to two factors: extensive pre-training on approximately 50 million synthetic data points~\cite{xue2025nmrmind} and the incorporation of 2D correlation spectra (e.g., HSQC, COSY), which enables the model to capture richer structural connectivity than 1D-only approaches. In contrast, earlier deep learning baselines exhibit distinct architectural limitations. {CLAMS} treats NMR spectra as dense 2D images, failing to extract effective features from sparse peaks (0.00\% sequence accuracy), while {NMR2Struct} suffers from a high rate of syntactically invalid outputs due to the lack of structural constraints during decoding.

For the search-based baseline, {NMR-Solver}, we explicitly report the performance of its \textit{Search} mode in the main comparison. We excluded the iterative optimization module for the full test set evaluation because its computational cost—averaging 8 minutes per sample—would necessitate over a month to process the entire test dataset. Consequently, we conducted the full optimization evaluation on a subset, the results of which are detailed in the Appendix~\ref{app:full_subset}. Even in Search-Only mode, NMR-Solver proves highly competitive (17.92\% Top-1 sequence accuracy) by leveraging a massive database of over 106 million compounds~\cite{jin2025nmr}. 

Regarding diffusion-based approaches, {DiffNMR} fails to generate exact matches (0.00\% accuracy) but captures some structural motifs. We evaluated {ChefNMR} in both zero-shot and fine-tuned settings. Motivated by its reported SOTA performance on synthetic benchmarks~\cite{hu2024accurate}, we first tested the zero-shot setting. However, consistent with limitations discussed by the authors~\cite{xiong2025atomic}, the substantial domain gap between synthetic and real-world data severely constrains its capability, resulting in near-zero sequence-level accuracy (0.02\%). Fine-tuning ChefNMR provides marginal improvements (reaching 1.93\% for ChefNMR-L), yet it remains significantly behind Transformer-based and search-based paradigms. We attribute this to the inefficient convergence of diffusion models, which often demand an excessive $\sim$10,000 epochs to reach stability~\cite{xiong2025atomic}.

\noindent\textbf{NMR2Func.} Given the scarcity of specialized deep learning approaches dedicated to functional group identification from NMR spectra, we established a comprehensive benchmark using standard neural architectures (CNN, MLP) and classical machine learning algorithms (XGBoost, Random Forest). Detailed hyperparameters and architectural configurations are provided in Appendix~\ref{nmr_func_tox}. As presented in Table~\ref{NMR2Func}, {XGBoost} demonstrates robust performance, achieving the highest Subset Accuracy (45.68\%) and Micro-F1 scores. However, the significant discrepancy between Micro- and Macro-metrics across all models highlights the severe \textit{class imbalance} inherent in this task, reflecting the long-tailed distribution characteristic of real-world chemical data. Despite this challenge, {NMR2Struct} yields competitive Macro-F1 scores (55.44\%), indicating that transformer-based backbones can effectively capture spectral features for diverse functional groups.

\noindent\textbf{NMR2Tox.} We evaluated toxicity prediction using the same set of general-purpose baselines (refer to Appendix~\ref{nmr_func_tox}). The results in Table~\ref{NMR2Tox} underscore the extreme difficulty of this task, which is compounded by two realistic constraints: (1) severe class imbalance, and (2) the \textit{low-resource} nature of the dataset (being a subset of the main corpus). While ensemble methods like {XGBoost} and {Random Forest} achieve high nominal accuracy ($>91\%$), this metric is misleading; their negligible recall scores indicate a failure to identify the minority toxic samples, a common pitfall in imbalanced, few-shot scenarios. In contrast, the {CNN} baseline achieves the state-of-the-art Macro-F1 (20.58\%) and Micro-F1 (26.42\%), significantly outperforming tree-based methods. This suggests that deep learning feature extractors offer better generalization for detecting subtle toxicity signals in this data-scarce, imbalanced regime.

\noindent\textbf{Mol2NMR.} Table~\ref{tab:nmr_comparison_refined} presents a comparative analysis of forward spectral simulation methods against {Mnova}, the commercial industry standard which employs a hybrid ensemble of algorithms including machine learning, HOSE-code, and Increments-based methods\cite{mnova}. All data-driven models are evaluated in a \textit{zero-shot} manner on experimental benchmarks.

Among open-source approaches, {NMRNet} demonstrates superior robustness, achieving 100\% chemical space coverage compared to the limited coverage of {DetaNet} (85.59\%). We attribute {NMRNet}'s success to its integration of 3D conformational information and extensive self-supervised geometric pre-training, which enables it to capture subtle stereochemical effects essential for high-fidelity spectral synthesis (surpassing {Mnova} in $^{13}$C Set Similarity: 0.9080 vs. 0.7268). In contrast, our architectural analysis reveals distinct limitations in {DetaNet}. While employing an advanced E(3)-equivariant tensor architecture, {DetaNet} relies on a fixed-vocabulary atomic embedding constrained by a pre-defined maximum atomic number (e.g., trained only on \textit{C, H, O, N, F} atom types). The observed incomplete coverage is primarily attributed to this design choice as well as the structural complexity inherent in natural products; the intricate heavy-atom scaffolds and high stereochemical density of these molecules often lead to computational convergence failures during inference for {DetaNet}.

Finally, we highlight a critical computational distinction between the top-performing open-source model, {NMRNet}, and the commercial standard {Mnova}. The hybrid algorithms of Mnova are predominantly CPU-bound. Consequently, computational latency increases significantly when analyzing complex natural products, where the combinatorial complexity of HOSE-code searches and rule-based increments struggles with intricate scaffolds. Conversely, {NMRNet} leverages deep learning architectures optimized for GPU acceleration. This allows for massively parallelized inference, enabling rapid spectral prediction that remains computationally efficient even for structurally complex molecules that typically bottleneck traditional CPU-based workflows.

\vspace{-3mm}
\subsection{Main Results}
\label{main}
\textbf{Data Leakage Analysis.} 
To rigorously evaluate the degree of data leakage, we conducted a quantitative comparison between the random and scaffold splits. Due to the prohibitive computational cost of calculating high-dimensional pairwise fingerprints across the full dataset, we performed this analysis on a representative subset of 1,000 samples randomly extracted from the test set. 
We computed the Maximum Tanimoto Similarity (MTS) for each sample against the training set, explicitly noting that instances with an MTS of $1.0$ frequently correspond to \textit{stereoisomers} (e.g., enantiomers) which share identical 2D graph connectivity but differ in 3D spatial arrangement, rather than exact duplicates. 
As visualized in Figure \ref{fig:split_analysis}, the random split (red) exhibits high structural redundancy with a mean similarity of $0.72$ and extensive coverage ($83.9\%$ at $\tau=0.5$). In stark contrast, the scaffold split (blue) induces a significant distributional shift towards lower similarity ($\Delta \approx 0.21$) and a sharp drop in coverage ($47.3\%$ at $\tau=0.5$), demonstrating that our benchmark effectively minimizes structural overlap.
\vspace{-3mm}
\subsection{Visualization}
\begin{figure}[t]
    \centering
    \includegraphics[width=0.86\linewidth]{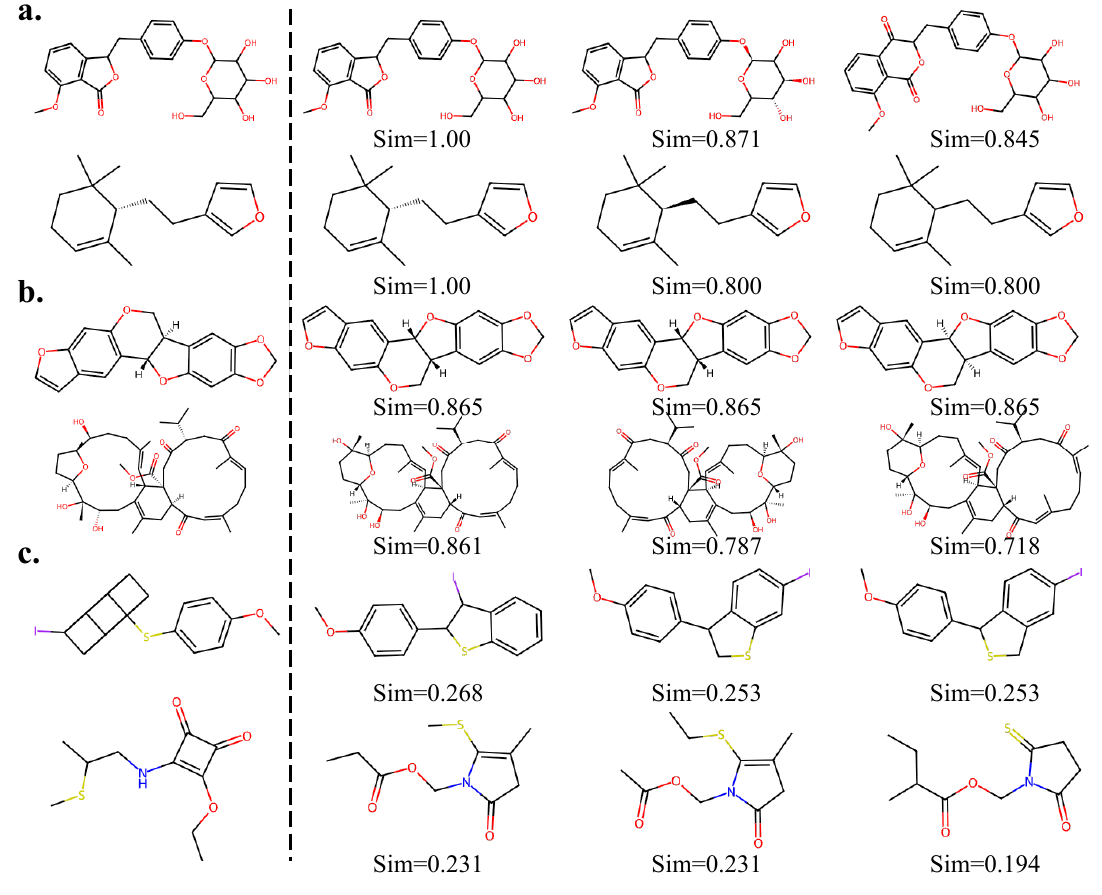}
    \caption{\textbf{Qualitative visualization of structural elucidation results.} We categorize the resultss into three representative scenarios: \textbf{(a)} \textit{Accurate Elucidation}. \textbf{(b)} \textit{High-Similarity Deviations}. \textbf{(c)} \textit{Low-Similarity Failures}. Note: ``Sim'' denotes the Tanimoto Similarity calculated using Morgan fingerprints.}
    \label{fig:vis_analysis}
\end{figure}

To provide deeper insights into the model's capabilities beyond aggregate metrics, we visualize representative structure elucidation results from best-performing deep learning model, {NMRMind}~\cite{xue2025nmrmind}, in Figure \ref{fig:vis_analysis}. The generated molecules are evaluated based on their Tanimoto Similarity to the ground truth using Morgan fingerprints.

\noindent\textbf{Accurate Elucidation (Panel a).} 
As shown in Figure \ref{fig:vis_analysis}(a), NMRMind demonstrates exceptional capability in reconstructing complex molecular scaffolds. In these successful cases, the model effectively aligns spectral peaks with the correct atom types and connectivity, achieving a perfect match (Sim $=$ 1.0).

\noindent\textbf{High-Similarity Deviations (Panel b).} 
Figure \ref{fig:vis_analysis}(b) illustrates NMRMind generates molecules with high fingerprint similarity (Sim $> 0.8$) but incorrect exact structures. In these instances, the model successfully identifies the dominant molecular scaffold (e.g., the fused ring systems) but may misplace specific functional group substitutions. This suggests that NMRMind has effectively learned the semantic mapping from spectra to substructures, even when it occasionally struggles with fine-grained atomic positioning.

\noindent\textbf{Low-Similarity Failures} 
In Figure \ref{fig:vis_analysis}(c), we observe failure cases characterized by low similarity scores. These errors occur in molecules with highly flexible macrocycles or unusual electronic environments where the spectral signals are ambiguous. This highlights the challenge of distinguishing structures with similar spectra.

\vspace{-3.5mm}
\section{Discussion}
\label{sec:discussion}

Our comprehensive benchmark reveals both the capabilities and the significant limitations of current AI methodologies in NMR spectral analysis. While Transformer-based models have established new state-of-the-art results, several critical challenges remain addressed to bridge the gap towards practical automation.
\noindent\textbf{Simulation-Reality Gap.}
A consistent trend across our experiments is the performance degradation attributable to the domain gap between synthetic training sets and real-world benchmarks. However, this does not negate the value of synthetic data; on the contrary, given the prohibitive cost of acquiring paired experimental datasets, high-fidelity simulation remains the cornerstone for scaling foundation models in this domain. Consequently, the field must pivot towards advanced adaptation paradigms specifically \textit{transfer learning} and \textit{test-time adaptation (TTA)}, enabling robust generalization, thereby effectively bridging the gap between in silico simulations and experimental reality.

\noindent\textbf{Peak-Atom Interpretability.}
Most current state-of-the-art methods operate as end-to-end black boxes, directly mapping spectral inputs to SMILES strings. While effective, this paradigm bypasses the fundamental chemical reasoning process: \textit{peak assignment} (i.e., mapping specific spectral peaks to specific atoms). This lack of explicit peak-atom alignment results in poor interpretability. Future architectures must move beyond simple sequence generation to incorporate explicit assignment mechanisms.

\noindent\textbf{Scarcity of 2D Data.}
Our results suggest that integrating 2D spectral constraints might have a significant impact on enhancing resolution power, particularly for complex isomers where 1D spectra suffer from severe peak overlap. However, a primary bottleneck lies in the scarcity of publicly available 2D NMR data relative to 1D datasets. Although 1D spectra are ubiquitous, they frequently lack explicit connectivity information, such as the direct Carbon-Hydrogen (C-H) correlations provided by Heteronuclear Single Quantum Coherence (HSQC) or the Proton-Proton (H-H) couplings revealed by Correlation Spectroscopy (COSY), which are essential for unambiguous structure determination. 

\vspace{-3mm}
\section{Conclusion}
In this work, we introduce \textbf{NMRGym}, the largest and first publicly released standardized benchmark derived from high-quality experimental NMR data, aiming to bridge the domain gap between synthetic and experimental spectra, standardizing future research and accelerating the deployment of \textit{AI4Spectrum}.





\bibliographystyle{ACM-Reference-Format}
\bibliography{sample-base}


\clearpage
\appendix

\section{Chemical Preliminaries}
\label{chemical}
Formally, we model the spectrum as a continuous function $x(\delta): \mathbb{R} \to \mathbb{R}$ over the chemical shift domain $\delta$. Disregarding spin-spin interactions, the signal is a superposition of $N$ independent resonance peaks:

\begin{equation}
    x(\delta) = \sum_{n=1}^{N} I_n \cdot \mathcal{L}(\delta; \mu_n, \lambda_n) + \xi(\delta),
    \label{eq:ideal_nmr}
\end{equation}

where $I_n$ and $\mu_n$ denote the intensity and chemical shift of the $n$-th nucleus, and $\xi(\delta)$ represents additive Gaussian noise. The spectral lineshape $\mathcal{L}$ typically follows a Lorentzian distribution with half-width $\lambda_n$:

\begin{equation}
    \mathcal{L}(\delta; \mu_n, \lambda_n) = \frac{1}{\pi} \frac{\lambda_n}{(\delta - \mu_n)^2 + \lambda_n^2}.
    \label{eq:lorentzian}
\end{equation}

In experimental settings, scalar coupling ($J$-coupling) introduces spin-spin interactions between neighboring nuclei, splitting the resonance signal into multiplets. The model generalizes to a nested summation over $K_n$ sub-peaks:
\begin{equation}
    x(\delta) = \sum_{n=1}^{N} \sum_{k=1}^{K_n} I_{n,k} \cdot \mathcal{L}(\delta; \mu_{n,k}, \lambda_{n}) + \xi(\delta),
    \label{eq:coupling_nmr}
\end{equation}

where the relative positions $\mu_{n,k}$ and intensities $I_{n,k}$ are governed by the coupling constants ($J$-values) and molecular topology.

Crucially, the accessibility of these spectral parameters varies significantly across data sources. High-fidelity datasets typically provide comprehensive annotations, including precise chemical shifts ($\mu$), peak intensities ($I$), and coupling constants ($J$).

\section{Implementation Details}
\label{nmr_func_tox}

We evaluated two categories of baselines: classic machine learning algorithms and deep neural networks. To ensure reproducibility, deterministic training was enforced with fixed random seeds (e.g., 42, 123, 456).
\subsection{Data Preprocessing}
All NMR spectra were transformed from discrete chemical shifts into continuous spectral representations via Gaussian broadening. We discretized $^{1}$H NMR spectra ($0-12$ ppm) and $^{13}$C NMR spectra ($0-220$ ppm) into 1,800 bins each. A Gaussian kernel ($\sigma = 0.05$ ppm) was applied to smooth the signals:
\begin{equation}
    y(x) = \sum_{i} \exp\left(-\frac{(x - s_i)^2}{2\sigma^2}\right),
\end{equation}
where $s_i$ denotes individual peak locations. The resulting spectra were min-max normalized to $[0, 1]$. For vector-based models (MLP, XGBoost, Random Forest), the two spectra were concatenated into a 3,600-dimensional vector. For 2D-based models (CNN), the spectra were reshaped into $60 \times 60$ dual-channel images.

\subsection{Classic Machine Learning Baselines}
\paragraph{XGBoost.} We trained an ensemble of independent binary classifiers for each functional group using the gradient boosting framework. Key hyperparameters included 200 boosting rounds, a maximum tree depth of 8, and a learning rate of 0.1. To mitigate class imbalance, we excluded extremely rare classes ($<10$ samples) and employed early stopping based on validation loss.

\paragraph{Random Forest.} We utilized a multi-output classification strategy where each forest comprised 200 trees with a maximum depth of 30. To address the skewed label distribution, we applied balanced class weighting (\texttt{class\_weight='balanced'}), which automatically adjusts weights inversely proportional to class frequencies.

\subsection{Deep Learning Baselines}
\paragraph{MLP}The MLP processed the flattened 3,600-dimensional spectral vectors through three fully connected layers ($3600 \to 1024 \to 512 \to 22$). Regularization included ReLU activations and dropout ($p=0.2$) after the first two layers.

\paragraph{CNN)}The CNN architecture consisted of four convolutional blocks followed by a three-layer fully connected head. Each block featured a $3 \times 3$ 2D convolution, batch normalization, ReLU activation, and pooling (max or adaptive average). Channel depth progressed as $1 \to 64 \to 128 \to 256 \to 512$. The classification head ($512 \to 1024 \to 512 \to 22$) included dropout ($p=0.3, 0.2$) to prevent overfitting.

\subsection{Training and Optimization}
\paragraph{Loss Function.} To counter severe class imbalance (positive rates $<1\% \sim 30\%$), we replaced standard cross-entropy with \textbf{Focal Loss} for all neural models. The loss is defined as:
\begin{equation}
    L_{FL} = -\sum_{i} [\alpha (1-p_i)^\gamma y_i \log(p_i) + (1-\alpha) p_i^\gamma (1-y_i) \log(1-p_i)].
\end{equation}
We set the focusing parameter $\gamma = 5.0$ and $\alpha = 0.25$ to down-weight easy negatives and emphasize hard, minority samples.

\paragraph{Optimization.} Deep models were optimized using Adam ($\text{lr} = 3 \times 10^{-6}$) with a step decay scheduler ($\gamma = 0.975$ per epoch). Training ran for up to 50 epochs with early stopping (patience = 6 epochs) and a batch size of 256.

\paragraph{Bias Initialization.} The final classification layer biases were initialized using empirical log-prior probabilities ($b_i = \log(p_i/(1-p_i))$) to align initial predictions with the marginal class distribution.

\subsection{Inference Strategy}
Instead of a fixed threshold (e.g., 0.5), we computed \textbf{optimal per-class thresholds} by maximizing the F1-score on the validation set. This adaptive thresholding strategy is critical for multi-label tasks with varying base rates.

\begin{table}[h]
    \centering
    \caption{Hyperparameter settings for baseline models.}
    \label{tab:hyperparams}
    \setlength{\tabcolsep}{8pt} 
    \begin{tabular}{l|c|c}
        \toprule
        \textbf{Method} & \textbf{Hyperparameter} & \textbf{Value} \\
        \midrule
        \multirow{3}{*}{General} & Gaussian $\sigma$ & 0.05 ppm \\
        & Spectral Bins ($^{1}$H / $^{13}$C) & 1800 / 1800 \\
        & Random Seeds & 42, 1337, 2024 \\
        \midrule
        \multirow{5}{*}{CNN / MLP} & Focal Loss ($\gamma$, $\alpha$) & 5.0, 0.25 \\
        & Learning Rate & $3 \times 10^{-6}$ \\
        & Batch Size & 256 \\
        & Optimizer & Adam \\
        & Dropout & $0.2 - 0.3$ \\
        \midrule
        \multirow{3}{*}{XGBoost} & n\_estimators & 200 \\
        & max\_depth & 8 \\
        & learning\_rate & 0.1 \\
        \midrule
        \multirow{3}{*}{Random Forest} & n\_estimators & 200 \\
        & max\_depth & 30 \\
        & class\_weight & balanced \\
        \bottomrule
    \end{tabular}
\end{table}

\section{Functional Group Definitions}
The definitions of functional groups are shown in Table~\ref{tab:functional_groups}.
\label{func_def}
\begin{table}
    \centering
    \caption{Definitions of the 20 functional groups labels included in the NMRGym.}
    \label{tab:functional_groups}
    \begin{tabularx}{\linewidth}{l >{\raggedright\arraybackslash\ttfamily}X}
        \toprule
         & \multicolumn{1}{l}{\textrm{Definition}} \\
        \midrule
        Alcohol & [OX2H] [CX4;!\$(C([OX2H])[O,S,\#7,\#15])] \\
        Carboxylic Acid & [CX3](=O)[OX2H1] \\
        Ester & [\#6][CX3](=O)[OX2H0][\#6] \\
        Ether & [OD2]([\#6])[\#6] \\
        Aldehyde & [CX3H1](=O)[\#6] \\
        Ketone & [\#6][CX3](=O)[\#6] \\
        Alkene & [CX3]=[CX3] \\
        Alkyne & [\$([CX2]\#C)] \\
        Benzene & c1ccccc1 \\
        Primary Amine & [NX3;H2;!\$(NC=[\! \#6]);!\$(NC\#[\! \#6])][\#6] \\
        Secondary Amine & [NH1,nH1] \\
        Tertiary Amine & [NH0,nH0] \\
        Amide & [NX3][CX3](=[OX1])[\#6] \\
        Cyano & [NX1]\#[CX2] \\
        Fluorine & [\#6][F] \\
        Bromine & [\#6][Br] \\
        Sulfonamide & [\#16X4]([NX3])(=[OX1])(=[OX1])[\#6] \\
        Sulfone & [\#16X4](=[OX1])(=[OX1])([\#6])[\#6] \\
        Sulfide & [\#16X2H0] \\[1ex]
        Phosphoric Acid & 
        [\$(P(=[OX1])([\$([OX2H]),\allowbreak\$([OX1-]),\allowbreak\$([OX2]P)])\allowbreak([\$([OX2H]),\allowbreak\$([OX1-]),\allowbreak\$([OX2]P)])\allowbreak[\$([OX2H]),\allowbreak\$([OX1-]),\allowbreak\$([OX2]P)]),\allowbreak
        \$([P+]([OX1-])([\$([OX2H]),\allowbreak\$([OX1-]),\allowbreak\$([OX2]P)])\allowbreak([\$([OX2H]),\allowbreak\$([OX1-]),\allowbreak\$([OX2]P)])\allowbreak[\$([OX2H]),\allowbreak\$([OX1-]),\allowbreak\$([OX2]P)])] \\
        \bottomrule
    \end{tabularx}
\end{table}
\section{Toxicity Label Definitions}
\label{tox_def}
The NMRGym toxicity subset includes 12 binary labels derived from the Tox21 Data Challenge. These labels represent experimental outcomes from quantitative high-throughput screening (qHTS) assays, categorized into Nuclear Receptor (NR) signaling pathways and Stress Response (SR) pathways. The definitions of functional groups are shown in Table~\ref{tab:tox_labels}.
\begin{table*}[htbp]
    \centering
    \caption{Definitions of the 12 toxicity labels included in the NMRGym subset.}
    \label{tab:tox_labels}
    \resizebox{0.6\linewidth}{!}{%
        \begin{tabular}{l|l|l}
            \hline
            \textbf{Label ID} & \textbf{Full Name} & \textbf{Biological Target / Mechanism} \\
            \hline
            \multicolumn{3}{l}{\textit{Nuclear Receptor (NR) Panel}} \\
            \hline
            NR-AR & Androgen Receptor & Agonism of the androgen receptor. \\
            NR-AR-LBD & Androgen Receptor LBD & Antagonism of the AR Ligand Binding Domain (luciferase). \\
            NR-AhR & Aryl Hydrocarbon Receptor & Activation of the aryl hydrocarbon receptor signaling. \\
            NR-Aromatase & Aromatase Enzyme & Inhibition of the aromatase enzyme (CYP19A1). \\
            NR-ER & Estrogen Receptor & Agonism of the estrogen receptor $\alpha$ (ER$\alpha$). \\
            NR-ER-LBD & Estrogen Receptor LBD & Antagonism of the ER Ligand Binding Domain. \\
            NR-PPAR-gamma & PPAR Gamma & Agonism of the Peroxisome Proliferator-Activated Receptor $\gamma$. \\
            \hline
            \multicolumn{3}{l}{\textit{Stress Response (SR) Panel}} \\
            \hline
            SR-ARE & Antioxidant Response Element & Activation of Nrf2 antioxidant pathway (oxidative stress). \\
            SR-ATAD5 & ATAD5 & Induction of genotoxicity (DNA damage response). \\
            SR-HSE & Heat Shock Response & Activation of heat shock factor response elements. \\
            SR-MMP & Mitochondrial Membrane Potential & Disruption of mitochondrial membrane potential. \\
            SR-p53 & p53 & Activation of the p53 DNA damage response pathway. \\
            \hline
        \end{tabular}%
    }
\end{table*}

\section{More Results of Structure Elucidation}
\subsection{Fingerprint Similarity Evaluation using Cosine Metric}
\begin{table*}[htbp]
\centering
\caption{Top-K Morgan, torsion, and atom-pair fingerprint similarity evaluation using cosine similarity.}
\resizebox{\linewidth}{!}{
\begin{tabular}{l c c c c c c c c c}
\toprule
& \multicolumn{3}{c}{\textbf{Morgan Sim (\% ↑)}} 
& \multicolumn{3}{c}{\textbf{Torsion Sim (\% ↑)}} 
& \multicolumn{3}{c}{\textbf{Atom-Pair Sim (\% ↑)}} \\
\cmidrule(lr){2-4}
\cmidrule(lr){5-7}
\cmidrule(lr){8-10}
& Top-1 & Top-5 & Top-10
& Top-1 & Top-5 & Top-10
& Top-1 & Top-5 & Top-10 \\
\midrule

\multicolumn{10}{l}{\textit{Search-based Methods}} \\
NMR-Solver(Search-Only)\cite{jin2025nmr} 
& 43.94 {\scriptsize $\pm$ 0.00} 
& 53.90 {\scriptsize $\pm$ 0.00}
& 57.75 {\scriptsize $\pm$ 0.00}
& 44.92 {\scriptsize $\pm$ 0.00}
& 56.27 {\scriptsize $\pm$ 0.00}
& 60.39 {\scriptsize $\pm$ 0.00}
& 60.38 {\scriptsize $\pm$ 0.00}
& 68.82 {\scriptsize $\pm$ 0.00}
& 71.77 {\scriptsize $\pm$ 0.00} \\
{\quad\itshape +Formula Condition}
& 45.68 {\scriptsize $\pm$ 0.00} 
& 47.41 {\scriptsize $\pm$ 0.00}
& 47.69 {\scriptsize $\pm$ 0.00}
& 46.08 {\scriptsize $\pm$ 0.00}
& 47.68 {\scriptsize $\pm$ 0.00}
& 47.92 {\scriptsize $\pm$ 0.00}
& 48.20 {\scriptsize $\pm$ 0.00}
& 49.30 {\scriptsize $\pm$ 0.00}
& 49.48 {\scriptsize $\pm$ 0.00} \\
\midrule

\multicolumn{10}{l}{\textit{Transformer-based Methods}} \\

CLAMS\cite{tan2025clams}
& 1.57 {\scriptsize $\pm$ 0.28}
& 4.80 {\scriptsize $\pm$ 0.88}
& 7.39 {\scriptsize $\pm$ 1.60}
& 0.59 {\scriptsize $\pm$ 0.19}
& 1.92 {\scriptsize $\pm$ 0.46}
& 3.07 {\scriptsize $\pm$ 0.67}
& 1.79 {\scriptsize $\pm$ 0.33}
& 5.87 {\scriptsize $\pm$ 1.19}
& 9.46 {\scriptsize $\pm$ 2.15} \\

NMRFormer\cite{nmrformer}
& 36.03 {\scriptsize $\pm$ 0.05}
& 52.45 {\scriptsize $\pm$ 0.06}
& 57.02 {\scriptsize $\pm$ 0.04}
& 37.14 {\scriptsize $\pm$ 0.04}
& 54.87 {\scriptsize $\pm$ 0.07}
& 59.88 {\scriptsize $\pm$ 0.06}
& 42.74 {\scriptsize $\pm$ 0.08}
& 61.89 {\scriptsize $\pm$ 0.05}
& 66.79 {\scriptsize $\pm$ 0.05} \\

NMR2Struct\cite{hu2024accurate}
& 48.93 {\scriptsize $\pm$ 0.66}
& 57.20 {\scriptsize $\pm$ 0.79}
& 61.01 {\scriptsize $\pm$ 0.75}
& 48.14 {\scriptsize $\pm$ 0.80}
& 58.43 {\scriptsize $\pm$ 0.83}
& 62.83 {\scriptsize $\pm$ 0.78}
& 62.17 {\scriptsize $\pm$ 0.54}
& 69.69 {\scriptsize $\pm$ 0.54}
& 72.63 {\scriptsize $\pm$ 0.48} \\

NMRMind\cite{xue2025nmrmind}
& 70.81 {\scriptsize $\pm$ 0.12}
& 76.77 {\scriptsize $\pm$ 0.08}
& 78.77 {\scriptsize $\pm$ 0.08}
& 71.68 {\scriptsize $\pm$ 0.10}
& 78.27 {\scriptsize $\pm$ 0.11}
& 80.46 {\scriptsize $\pm$ 0.09}
& 79.42 {\scriptsize $\pm$ 0.04}
& 84.20 {\scriptsize $\pm$ 0.15}
& 85.68 {\scriptsize $\pm$ 0.09} \\

{\quad\itshape +Formula Condition}
& \textbf{74.09 {\scriptsize $\pm$ 0.03}}
& \textbf{80.53 {\scriptsize $\pm$ 0.06}}
& \textbf{82.39 {\scriptsize $\pm$ 0.07}}
& \textbf{75.86 {\scriptsize $\pm$ 0.04}}
& \textbf{82.70 {\scriptsize $\pm$ 0.05}}
& \textbf{84.64 {\scriptsize $\pm$ 0.06}}
& \textbf{82.84 {\scriptsize $\pm$ 0.01}}
& \textbf{87.96 {\scriptsize $\pm$ 0.04}}
& \textbf{89.23 {\scriptsize $\pm$ 0.03}} \\

\midrule

\multicolumn{10}{l}{\textit{Diffusion-based Methods}} \\
DiffNMR\cite{yang2025diffnmr}
& 16.14 {\scriptsize $\pm$ 0.02}
& 23.09 {\scriptsize $\pm$ 0.01}
& 25.12 {\scriptsize $\pm$ 0.02}
& 11.58 {\scriptsize $\pm$ 0.01}
& 21.08 {\scriptsize $\pm$ 0.02}
& 24.41 {\scriptsize $\pm$ 0.01}
& 32.09 {\scriptsize $\pm$ 0.02}
& 43.30 {\scriptsize $\pm$ 0.01}
& 45.76 {\scriptsize $\pm$ 0.02} \\

ChefNMR-S\cite{xiong2025atomic}
& 4.70  {\scriptsize $\pm$ 0.04}
& 12.35 {\scriptsize $\pm$ 0.03}
& 15.83 {\scriptsize $\pm$ 0.04}
& 4.07  {\scriptsize $\pm$ 0.04}
& 11.74 {\scriptsize $\pm$ 0.06}
& 15.68 {\scriptsize $\pm$ 0.05}
& 10.41 {\scriptsize $\pm$ 0.07}
& 26.90 {\scriptsize $\pm$ 0.08}
& 33.97 {\scriptsize $\pm$ 0.14} \\

ChefNMR-S(Finetune)\cite{xiong2025atomic}
& 23.39 {\scriptsize $\pm$ 0.09} 
& 44.72 {\scriptsize $\pm$ 0.13}
& 51.13 {\scriptsize $\pm$ 0.18}
& 24.67 {\scriptsize $\pm$ 0.09}
& 47.83 {\scriptsize $\pm$ 0.18}
& 55.00 {\scriptsize $\pm$ 0.20}
& 32.80 {\scriptsize $\pm$ 0.11}
& 60.50 {\scriptsize $\pm$ 0.16}
& 67.61 {\scriptsize $\pm$ 0.21} \\

ChefNMR-L\cite{xiong2025atomic}
& 5.55  {\scriptsize $\pm$ 0.03}
& 14.24 {\scriptsize $\pm$ 0.04}
& 18.04 {\scriptsize $\pm$ 0.02}
& 4.89  {\scriptsize $\pm$ 0.02}
& 13.62 {\scriptsize $\pm$ 0.04}
& 17.98 {\scriptsize $\pm$ 0.02}
& 12.05 {\scriptsize $\pm$ 0.01}
& 30.29 {\scriptsize $\pm$ 0.01}
& 37.74 {\scriptsize $\pm$ 0.01} \\

ChefNMR-L(Finetune)\cite{xiong2025atomic}
& 27.52 {\scriptsize $\pm$ 0.20} 
& 49.08 {\scriptsize $\pm$ 0.06}
& 54.80 {\scriptsize $\pm$ 0.04}
& 29.08 {\scriptsize $\pm$ 0.11}
& 52.54 {\scriptsize $\pm$ 0.07}
& 58.89 {\scriptsize $\pm$ 0.05}
& 38.29 {\scriptsize $\pm$ 0.29}
& 65.13 {\scriptsize $\pm$ 0.16}
& 70.86 {\scriptsize $\pm$ 0.05} \\

\bottomrule
\end{tabular}
}
\end{table*}

\subsection{Performance of NMR-Solver (Search + Optimization) on a Subset}
\label{app:full_subset}
\begin{figure*}
    \centering
    \includegraphics[width=0.8\linewidth]{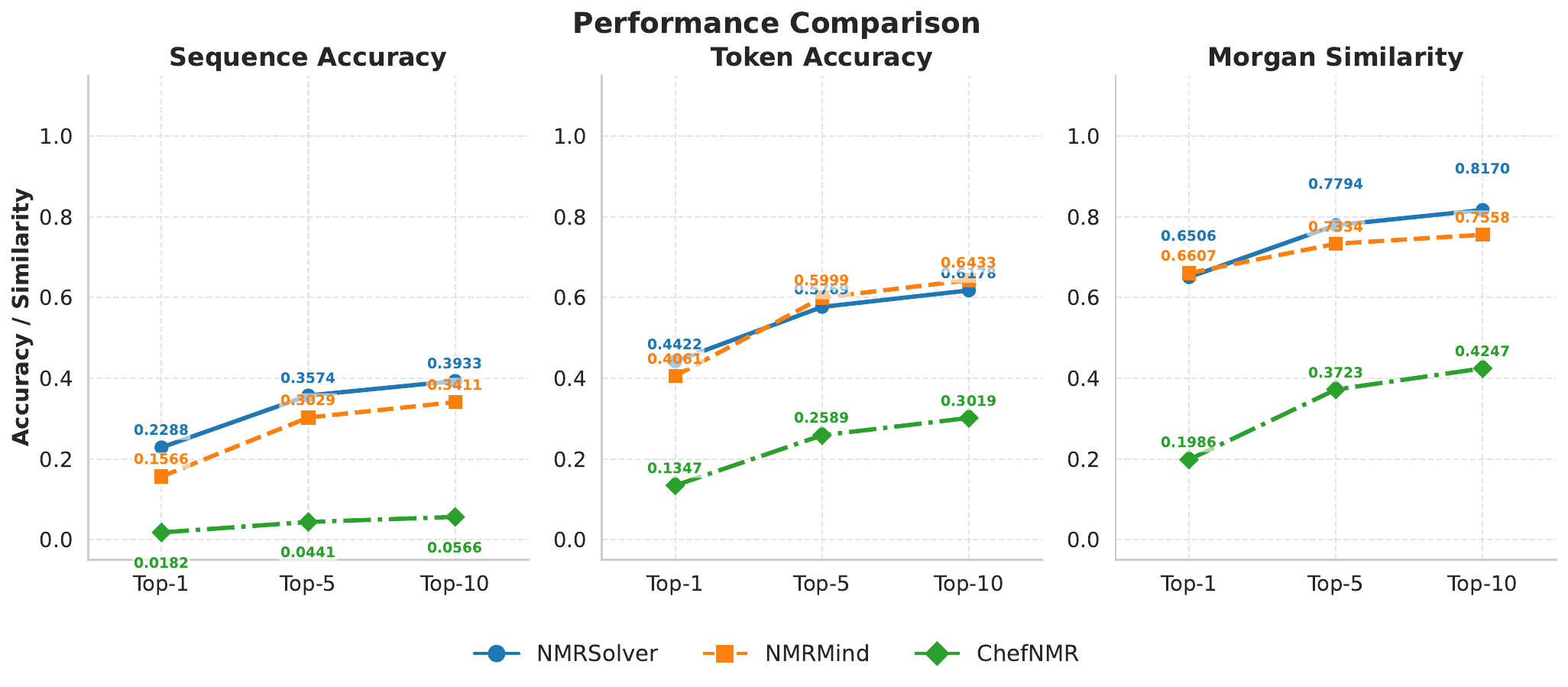}
    \caption{{Performance comparison on the hard subset.} Evaluation of Sequence Accuracy, Token Accuracy, and Morgan Tanimoto Similarity across Top-1, Top-5, and Top-10 rankings. {NMRSolver} (utilizing the full pipeline with the optimization module) is compared against {NMRMind} and the baseline {ChefNMR}. The results demonstrate the efficacy of the optimization strategy in refining structural predictions.}
    \label{fig:subset_performance}
\end{figure*}
Given that the {NMRSolver} full pipeline entails computationally intensive 3D conformer generation and combinatorial optimization—which would require over one month to process the entire test set—we curated a subset of 6,329 samples to benchmark against state-of-the-art transformer-based and diffusion-based methods.

As illustrated in Figure~\ref{fig:subset_performance}, {NMRSolver (full pipeline with the optimization module)} demonstrates a significant performance advantage over the baseline methods, particularly in Sequence-Level Accuracy, which serves as the strictest metric for structure elucidation.
Most notably, {NMRSolver} achieves a Top-1 Sequence Accuracy of 22.88\%, outperforming {NMRMind} (15.66\%) by a substantial margin of over 7 absolute percentage points (a relative improvement of $\sim$46\%). Interestingly, while {NMRMind} maintains competitive performance in Token Level Accuracy (e.g., Top-5: 60.0\% vs. {NMRSolver}'s 57.7\%), it struggles to convert these correctly predicted substructures into the exact molecular structure.

This discrepancy highlights the critical contribution of our proposed \textit{Optimization Module}. By refining the structural candidates, the optimization process effectively corrects topological errors that pure sequence generation models (like {NMRMind}) fail to resolve. Furthermore, in terms of structural similarity, {NMRSolver} consistently retrieves candidates with higher chemical fidelity in the Top-5 and Top-10 rankings (Morgan Similarity: 0.78 and 0.82 vs. {NMRMind}'s 0.73 and 0.76), further validating the feasibility and robustness of the optimization-enhanced pipeline.
\end{document}